\documentclass[iop,apj,preprint]{emulateapj}

\usepackage[]{amsmath}
\usepackage{apjfonts}
\usepackage[dvipsnames]{xcolor}
\usepackage{bm}
\usepackage{graphicx}
\usepackage{ulem}
\usepackage{lipsum}

\shorttitle{Intra-cluster Light in the HFF Clusters}
\shortauthors{Morishita et al.}

\usepackage{color}			      
\definecolor{midgray}{gray}{0.4}		
\definecolor{orange}{rgb}{1,0.5,0}    
\definecolor{leablue}{rgb}{0,0,0}
\def\bfl{\color{leablue}}
\def\bfb{\color{blue}}

\def\nsys{500}

\definecolor{ao}{rgb}{0.0, 0.0, 1.0}
\usepackage[colorlinks=true, breaklinks=true, citecolor=ao, filecolor=ao, linkcolor=ao, urlcolor=ao]{hyperref}        
\usepackage{scrextend}
\deffootnote[0.5em]{0em}{1em}{\textsuperscript{\thefootnotemark}\,}

\usepackage{etoolbox}
\makeatletter
\patchcmd{\NAT@citex}
  {\@citea\NAT@hyper@{\NAT@nmfmt{\NAT@nm}\NAT@date}}
  {\@citea\NAT@nmfmt{\NAT@nm}\NAT@hyper@{\NAT@date}}
  {}
  {}
\patchcmd{\NAT@citex}
  {\@citea\NAT@hyper@{%
     \NAT@nmfmt{\NAT@nm}%
     \hyper@natlinkbreak{\NAT@aysep\NAT@spacechar}{\@citeb\@extra@b@citeb}%
     \NAT@date}}
  {\@citea\NAT@nmfmt{\NAT@nm}%
   \NAT@aysep\NAT@spacechar%
   \NAT@hyper@{\NAT@date}}
  {}
  {}
\patchcmd{\NAT@citex}
  {\@citea\NAT@hyper@{%
     \NAT@nmfmt{\NAT@nm}%
     \hyper@natlinkbreak{\NAT@spacechar\NAT@@open\if*#1*\else#1\NAT@spacechar\fi}%
       {\@citeb\@extra@b@citeb}%
     \NAT@date}}
  {\@citea\NAT@nmfmt{\NAT@nm}%
   \NAT@spacechar\NAT@@open\if*#1*\else#1\NAT@spacechar\fi%
   \NAT@hyper@{\NAT@date}}
  {}
  {}
\makeatother

\newcommand{\simgt}{\,\rlap{\lower 3.5 pt \hbox{$\mathchar \sim$}} \raise
1pt \hbox {$>$}\,}
\newcommand{\simlt}{\,\rlap{\lower 3.5 pt \hbox{$\mathchar \sim$}} \raise
1pt \hbox {$<$}\,}

\newcommand{\Msun}{M_{\odot}}
\newcommand{\logm}{\log M_*/\Msun}
\newcommand{\lsig}{\log \Sigma_*/\,\Msun\,{\rm kpc}^{-2}}

\newcommand{\kms}{{\rm km~s^{-1}}}

\newcommand{\hst}{{\it HST}}
\newcommand{\galfit}{{\ttfamily GALFIT}}
\newcommand{\sext}{{\ttfamily SExtractor}}

\newcommand{\fast}{{\ttfamily FAST}}
\newcommand{\galaxev}{{\ttfamily GALAXEV}}

\begin{document}
\slugcomment{Accepted for Publication in The Astrophysical Journal}
\title{Characterizing Intra-cluster Light in the Hubble Frontier Fields}
\author{
Takahiro~Morishita$^{1,2,3}$, 
Louis~E.~Abramson$^{1}$, 
Tommaso~Treu$^{1}$,
Kasper~B.~Schmidt$^{4}$,
Benedetta~Vulcani$^{5}$,
Xin~Wang$^{1}$
}
\altaffiltext{1}{Department of Physics and Astronomy, University of California, Los Angeles, CA 90095-1547, USA}
\altaffiltext{2}{Astronomical Institute, Tohoku University, Aramaki, Aoba, Sendai 980-8578, Japan}
\altaffiltext{3}{Institute for International Advanced Research and Education, Tohoku University, Aramaki, Aoba, Sendai 980-8578, Japan}
\altaffiltext{4}{Leibniz-Institut f\"ur Astrophysik Potsdam (AIP), An der Sternwarte 16, D-14482 Potsdam, Germany}
\altaffiltext{5}{School of Physics, The University of Melbourne, VIC 3010, Australia}
\email{mtaka@astro.ucla.edu}

\begin{abstract}
We investigate the intra-cluster light (ICL) in the 6\,Hubble Frontier Field clusters at $0.3<z<0.6$.
We employ a new method, which is free from any functional form of the ICL profile, and exploit the unprecedented depth of this {\it Hubble Space Telescope} imaging to map the ICL's diffuse light out to clustrocentric radii $R\sim300$\,kpc ($\mu_{\rm ICL}\sim27$\,mag\,arcsec$^{-2}$).
From these maps, we construct radial color and stellar mass profiles via SED fitting and find clear negative color gradients in all systems with increasing distance from the Brightest Cluster Galaxy (BCG). 
While this implies older/more metal rich stellar components in the inner part of the ICL, we find the ICL mostly consists of a $\simlt2$\,Gyr population, and plausibly originated with $\logm\simlt10$ cluster galaxies.
Further, we find 10--15\% of the ICL's mass at large radii ($\simgt150$\,kpc) lies in a younger/bluer stellar population ($\sim1$\,Gyr), a phenomenon not seen in local samples.
We attribute this light to the higher fraction of starforming/(post-)starburst galaxies in clusters at $z\sim0.5$.
Ultimately, we find the ICL's total mass to be $\log M_{\rm *}^{\rm ICL}/\,M_\odot\sim11$--12, constituting 5\%-20\,\% of the clusters' total stellar mass, or about a half of the value at $z\sim0$.
The above implies distinct formation histories for the ICL and BCGs/other massive cluster galaxies; i.e. the ICL at this epoch is still being constructed rapidly ($\sim40\,M_\odot$\,yr$^{-1}$), while the BCGs have mostly completed their evolution.
To be consistent with the ICL measurements of local massive clusters, such as the Virgo, our data suggest mass acquisition mainly from quiescent cluster galaxies is the principal source of ICL material in the subsequent $\sim$5 Gyr of cosmic time.
\end{abstract}

\section{Introduction}
Intra-cluster light (ICL) is starlight that fills the intergalactic space in dense galaxy environments. First proposed by \citet{zwicky37}, it is unique to galaxy groups and clusters, suggesting that its formation process is related to environment-specific phenomena that may also influence galaxies. Thus, understanding the origin and evolution of the ICL may aid our understanding of galaxy evolution.

Numerical calculations suggest that the ICL in massive clusters ($\log M_{\rm 500}/M_\odot\sim15$) formed {\bfl from stars stripped in galaxy/galaxy and galaxy/intracluster medium interactions at $z<1$ \citep{larson80, nipoti03, mcpartland16}} when massive cluster members (including brightest cluster galaxies; BCGs) had nearly completed their stellar mass accumulation \citep{murante07, collins09, contini14, burke15}. 
However, observational constraints on {\bfl the ICL's} origin and evolution have been limited due to {\bfl its} low surface brightness \citep[][]{zaritsky04, zibetti05, gonzalez07, toledo11, guennou12, presotto14}, and difficulties introduced by the fact that other galaxies are, by definition, embedded within it.

In this paper, we use deep {\it Hubble Space Telescope} ($HST$) multi-band imaging from the {\it Hubble Frontier Fields} \citep[HFF;][]{lotz17} to overcome these obstacles and dissect the ICL in {\bfl six} clusters {\bfl at $z\sim0.5$, an epoch when it is rapidly assembling.
We develop a new method to reconstruct the ICL {\bfl that entails no assumptions about its shape, but relies instead our more robust understanding} of individual galaxy light profiles.
{\bfl From these maps, we first infer the ICL's radial color and stellar population gradients at clustrocentric radii $R\lesssim300$\,kpc, and from there its formation and growth history to $z=0$. Finally, we attempt to quantitatively link the ICL to its probable cluster galaxy progenitors/sources using the above combined with a complete galaxy census ($\logm \gtrsim 7.8$)} from \citet[][hereafter M17]{morishita17}.

\begin{deluxetable}{lccc}\tablecolumns{4}
\tablecaption{Background Statistics}
\tablehead{
\colhead{Cluster} & \colhead{Filter} & \colhead{$\mu$\tablenotemark{\bfb a}} & \colhead{$\sigma$\tablenotemark{\bfb a}}\\
\colhead{} & \colhead{} & \colhead{(count s$^{-1}$ pix$^{-1}$)} & \colhead{(count s$^{-1}$ pix$^{-1}$)}}
\startdata
{\sc Abell2744} & F160W & 6.6633e-04 & 9.8069e-04\\  
 & F140W & 4.6230e-04 & 8.9982e-04\\
 & F125W & 5.5142e-04 & 8.4289e-04\\
 & F105W & 5.0786e-04 & 5.7572e-04\\
 & F814W & 3.3045e-05 & 4.3519e-04\\
 & F606W & -8.4930e-05 & 3.3544e-04\\
 & F435W & -2.5166e-04 & 3.5891e-04\\
{\sc Macs0416} & F160W & 1.3259e-04 & 8.2433e-04\\
 & F140W & 7.9862e-05 & 8.5225e-04\\
 & F125W & 1.1805e-04 & 7.4375e-04\\
 & F105W & 9.9649e-05 & 5.4596e-04\\
 & F814W & 5.0604e-04 & 3.8697e-04\\
 & F606W & 4.3288e-04 & 3.0466e-04\\
 & F435W & 7.7018e-04 & 3.4692e-04 \\
{\sc Macs0717} & F160W & 1.5467e-04 & 1.0552e-03 \\
 & F140W & 2.2186e-04 & 9.5267e-04 \\
 & F125W & 2.3522e-04 & 9.1900e-04 \\
 & F105W & 2.1276e-04 & 5.6932e-04 \\
 & F814W & 2.7187e-04 & 3.4344e-04 \\
 & F606W & 5.2020e-05 & 3.6513e-04 \\
 & F435W & 1.1210e-04 & 4.4185e-04 \\
{\sc Macs1149} & F160W & 5.4298e-04 & 1.2053e-03 \\
 & F140W & 5.2340e-04 & 1.2435e-03 \\
 & F125W & 4.4334e-04 & 1.1364e-03 \\
 & F105W & 4.1998e-04 & 6.1400e-04 \\
 & F814W & 8.4780e-05 & 5.1403e-04 \\
 & F606W & -8.0179e-05 & 3.1921e-04 \\
 & F435W & 1.7872e-04 & 5.7851e-04 \\
{\sc Abell0370}\tablenotemark{\bfb b} & F160W & -7.0873e-04 & 1.0893e-03 \\
 & F140W & -5.0910e-04 & 9.7446e-04 \\
 & F125W & -4.4832e-04 & 8.9771e-04 \\
 & F105W & -5.0517e-04 & 5.0632e-04 \\
 & F814W & -5.4013e-04 & 4.6356e-04 \\
 & F606W & -3.2906e-04 & 3.8140e-04 \\
 & F435W & 1.2233e-04 & 5.5065e-04 \\
{\sc AbellS1063} & F160W & 7.4988e-04 & 1.0244e-03 \\
 & F140W & 6.7013e-04 & 9.2828e-04 \\
 & F125W & 6.6327e-04 & 8.8715e-04 \\
 & F105W & 4.6818e-04 & 5.6195e-04 \\
 & F814W & 1.1040e-04 & 4.9730e-04 \\
 & F606W & 1.3584e-04 & 3.5992e-04\\
 & F435W & 6.9163e-04 & 4.6383e-04
\enddata
\label{tab:bg} \tablenotetext{a}{Gaussian fit to pixels in at $R>1200$\,pix from BCGs.} \tablenotetext{b}{The background in the HFF Abell0370 images appear to be over-subtracted by the HFF pipeline.}
\tablecomments{It is noted that the original sky is already subtracted in the \hst\ pipeline, and the estimated background values here are the residual of the subtraction process. 
}
\end{deluxetable}

\begin{deluxetable*}{lccccccc} \tablecolumns{8}
\tablewidth{0pt} 
\tabletypesize{\scriptsize} \tabcolsep=0.0cm
\tablecaption{Brightest Cluster Galaxies Information}
\tablehead{
\colhead{ID \tablenotemark{\bfb a}} & \colhead{RA} & \colhead{DEC} & \colhead{$\log M_*^{\rm cor}$ \tablenotemark{\bfb b}} & \colhead{$m_{\rm GALFIT}$ \tablenotemark{\bfb c}}  & \colhead{$q$ \tablenotemark{\bfb c}} & \colhead{PA \tablenotemark{\bfb c}} & \colhead{$m_{\rm AUTO}$ \tablenotemark{\bfb d}}\\
\colhead{} & \colhead{(J2000)} & \colhead{(J2000)} & \colhead{($\log M_\odot$)} & \colhead{(mag)} & \colhead{} & \colhead{(degree)} & \colhead{(mag)}}
\startdata
{01-1076} & 03:35:10.53 & -30:24:00.63 & 11.51 & 15.81 & 0.98 & -26.22 &16.81\\
{02-593}   & 64:02:17.10 & -24:04:02.96 & 11.38  & 16.71 & 0.77 & 55.40 &17.06\\
{03-1186} & 109:23:53.69 & 37:44:44.65 & 11.72 & 16.93 & 0.92 & 31.92 &17.93\\
{04-1310} & 177:23:55.47 & 22:23:54.70 & 11.60 & 16.94 & 0.75 & -55.13 &17.94\\
{05-696} & 39:58:10.91 & -01:34:18.80 & 11.75 & 15.80 & 0.83 & 49.20 &16.80\\
{06-914} & 342:10:59.59 & -44:31:51.16 & 11.60 & 15.29 & 0.75 & 51.74 &16.29
\enddata
\label{tab:bcg}
\tablenotetext{a}{F160W-selected Brightest Cluster Galaxies (BCGs).}
\tablenotetext{b}{Stellar mass corrected to the \galfit\ magnitude (Eq.~\ref{eq:mcor}).}
\tablenotetext{c}{\galfit\ structural parameters (magnitude, axis ratio, position angle).}
\tablenotetext{d}{\sext\ F160W-band auto magnitude from which initial BCG stellar masses are derived.}
\end{deluxetable*}

\begin{figure*}[t!]
\begin{center}
\includegraphics[width = 0.4\textwidth]{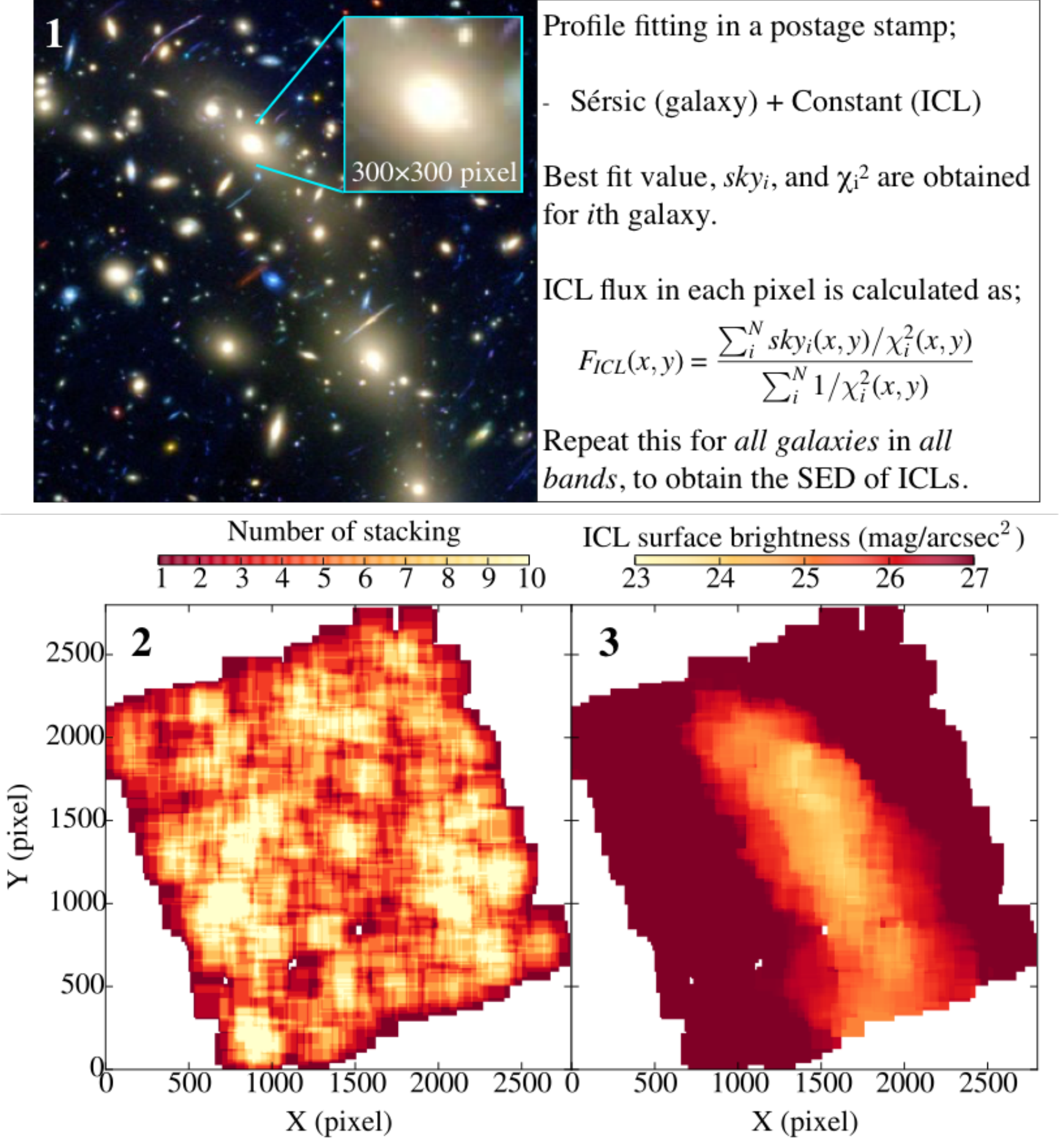}
\includegraphics[width = 0.45\textwidth, trim = -0.5cm 0cm 0.5cm 0cm]{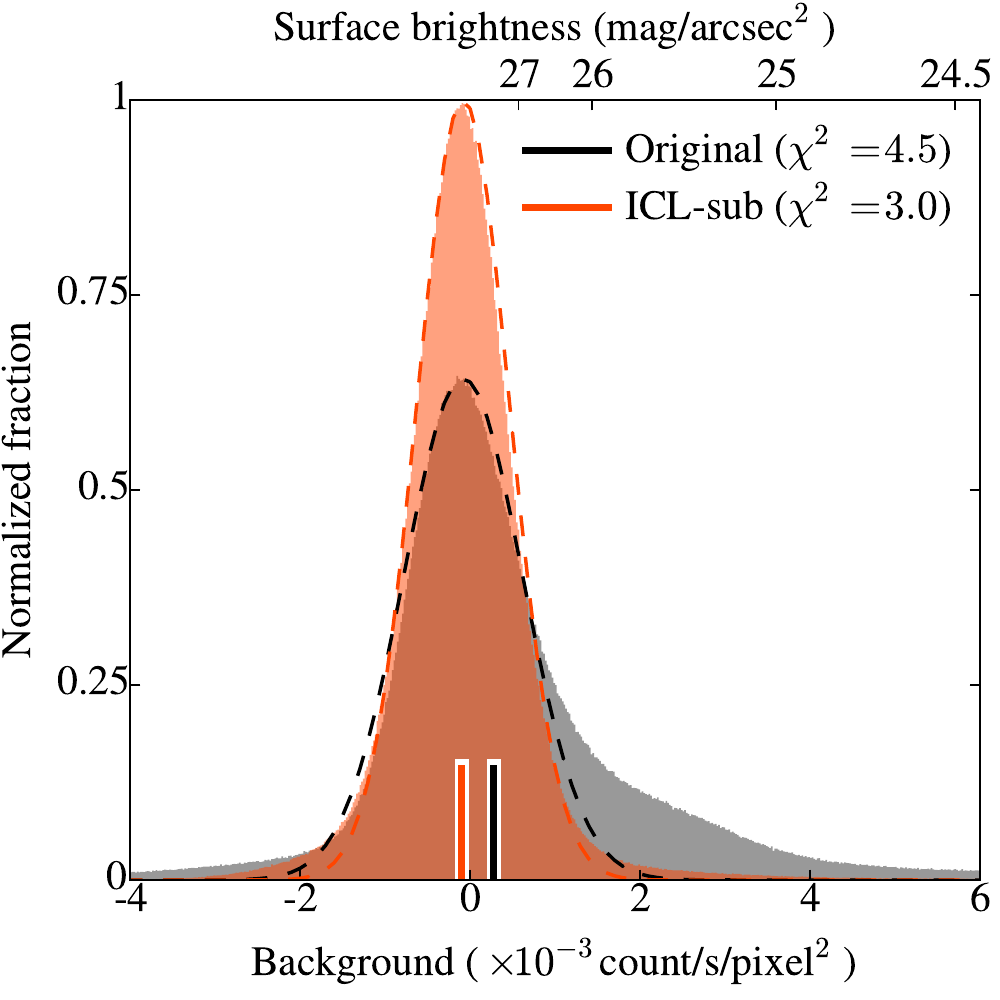}
\caption{
{\it Left:} Schematic view of ICL-subtraction method:
We fit the galaxy light profile and local sky background for each target galaxy in M17 (1), and stack the modeled sky in overlapped pixels (the number of overlap in each pixel is shown in 2), and stack the sky over the FoV of WFC3 ($\sim300$\,kpc) to construct the ICL map (3).
The stacking is weighted by the inverse of $\chi_{\rm GALFIT}^2$ values of the light profile fitting (Equation~\ref{eq:chi}).
{\it Right:} Comparison of two histograms for the pixel values in ICL-subtracted (red shaded region) and original (black region) mosaics of one example cluster (MACS0416/F160W-band).
Each histogram consists of pixels unmasked by the same \sext's detection map for both mosaics, and therefore the contribution from galaxies is suppressed.
An excess of positive pixel is observed in the original mosaic (i.e. ICL+light from undetected galaxies; see Section~\ref{ssec:und}), while the distribution is more symmetric for the ICL-subtracted image, as well as the medians of distributions indicate (bars of same colors as the distributions).
The fit with a gaussian for each distribution (dashed lines of same colors) and its goodness ($\chi_{\rm fit}^2$) are shown.
}
\label{fig:method}
\end{center}
\end{figure*}

Below, we assume  a \citet{chabrier03} initial mass function, $\Omega_m=0.3$, $\Omega_\Lambda=0.7$, and $H_0=70\,\kms\, {\rm Mpc}^{-1}$. Magnitudes are in the AB system \citep{oke83, fukugita96}. 

\section{Data}\label{sec:data}
We analyze the central (CLS) pointings covering the six HFF clusters: Abell2744, MACS0416, MACS0717, MACS1149, Abell370, and Abell\,S1063 (a.k.a. RXCJ2248). 
Imaging spans ACS F435/606/814W through WFC3IR F105/125/140/160W to a limiting F160W magnitude of $m_{160}\sim28.5$ \citep[][]{lotz17}. 
All imaging is PSF-matched to F160W resolution.
We base our comparison between the ICL and cluster galaxies properties (e.g., Section~\ref{ssec:col}) on the catalog published by M17.
While we briefly clarify in the following sections, more detailed information can be found in M17.


\begin{figure}
\begin{center}
\includegraphics[width=0.48\linewidth, trim= 0.07cm 0.07cm 0.07cm 0.07cm]{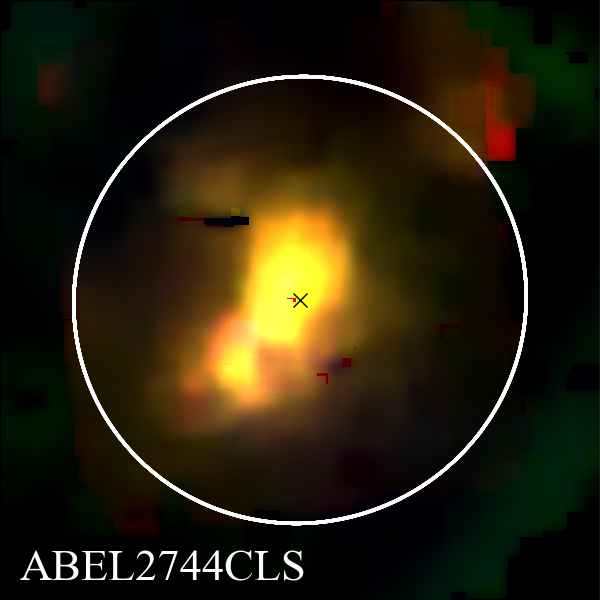}
\includegraphics[width=0.48\linewidth, trim= 0.07cm 0.07cm 0.07cm 0.07cm]{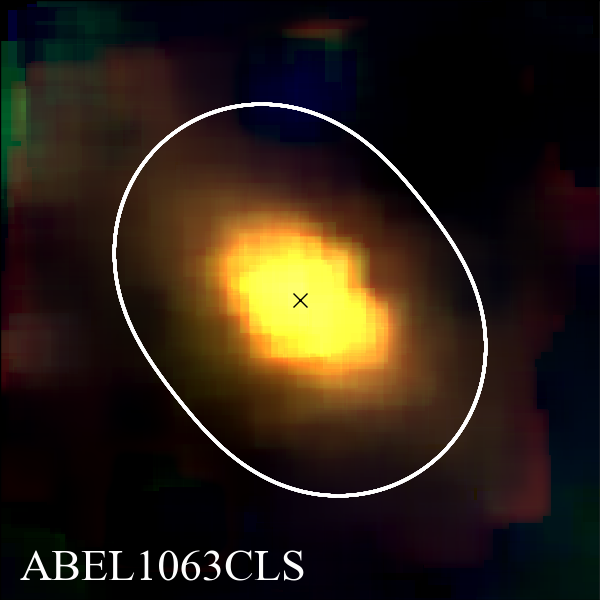}
\includegraphics[width=0.48\linewidth, trim= 0.07cm 0.07cm 0.07cm 0.07cm]{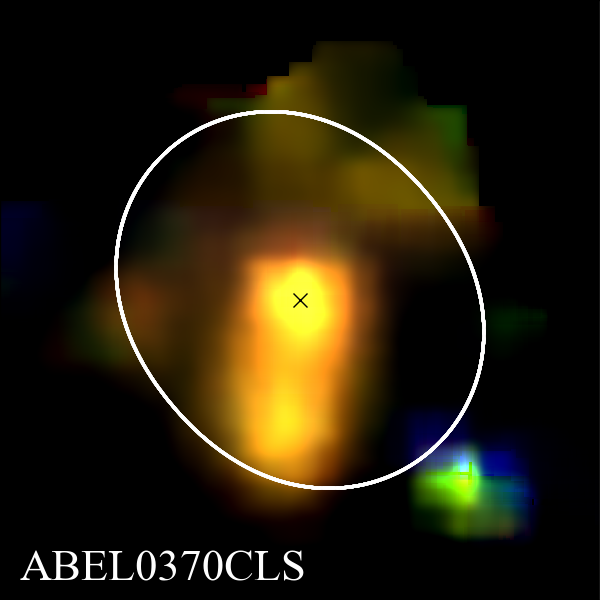}
\includegraphics[width=0.48\linewidth, trim= 0.07cm 0.07cm 0.07cm 0.07cm]{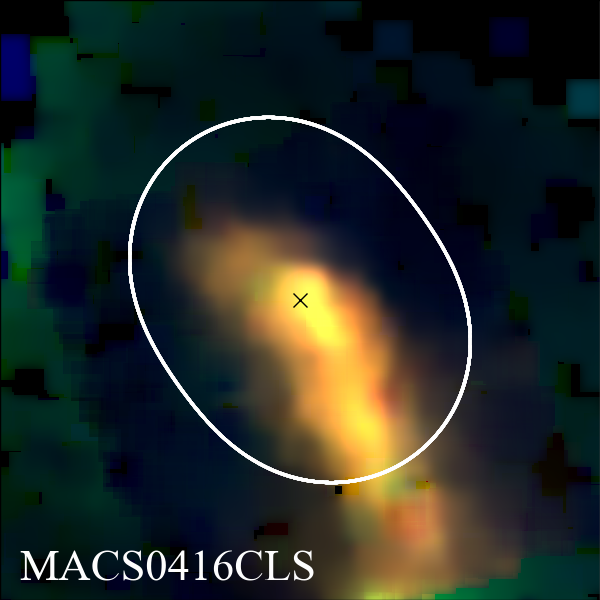}
\includegraphics[width=0.48\linewidth, trim= 0.07cm 0.07cm 0.07cm 0.07cm]{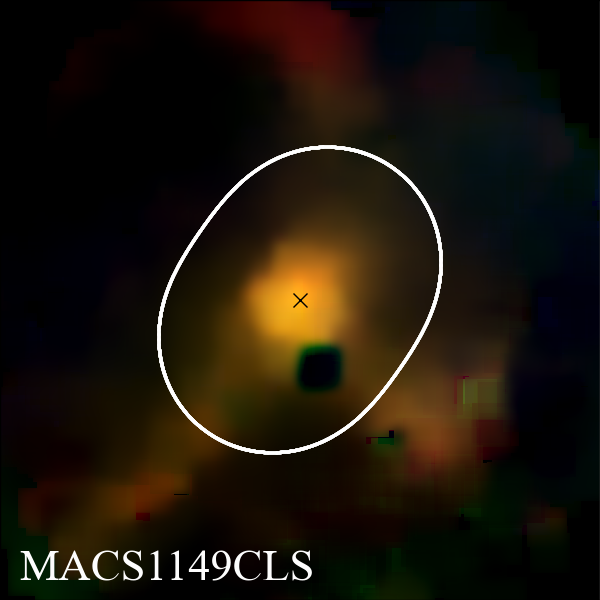}
\includegraphics[width=0.48\linewidth, trim= 0.07cm 0.07cm 0.07cm 0.07cm]{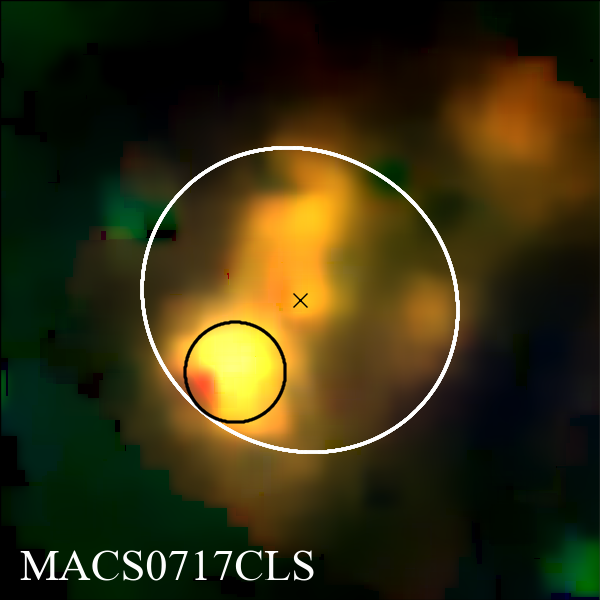}
\caption{
ICL maps in the HFF clusters. Images span $2800\times2800$\,pix (comparable to the WFC3 FoV), and colors are F435+606W (B), F814+105W (G), F125+140+160W (R).
White ellipses denote $R=300$\,kpc from the BCG (marked by crosses) with position angles and axis ratios reflecting those of the BCG.
Masked pixels near a bright foreground galaxy towards MACS0717 are outlined by a black circle (see Section~\ref{ssec:bcg}).
}
\label{fig:rgb}
\end{center}
\end{figure}

\section{ICL Reconstruction via Galaxy Fitting}\label{sec:method}

Figure~\ref{fig:method} shows a schematic of our ICL reconstruction method.
We detect all sources on the composite image using \sext\ \citep{bertin96}.
For those where structural fitting is reliable---$m_{160}<26$ ($\sim$400--900 sources per cluster)---we fit single S\'ersic profiles using \galfit\ \citep{peng02} in $300\times300$ pixel ($\sim$$100\times100$ kpc) postage stamps as in M17.
Neighboring galaxies brighter than the target are fit simultaneously; fainter sources are masked.

The key output of this process is the {\it local} (constant) background level \galfit\ finds for each stamp. This quantity forms the basis of our ICL flux estimate, but is also sensitive to {\it global} backgrounds from non-ICL sources (e.g., zodiacal light, earthshine).
Such contributions are small in the NIR, but become meaningful in blue bandpasses where the ICL is faint.
We estimate and subtract these pedestals} before SED fitting (Section~\ref{ssec:bg}).

After finding the local backgrounds for all $m_{\rm 160}<26$ sources, we build a map of the ICL by laying down stamps set to these values.
Stamps generally overlap, so we calculate the representative ICL value in each pixel, $F_{\rm ICL}(x,y)$, using their $\chi^{2}$-weighted mean:
\begin{equation}\label{eq:chi}
		F_{\rm ICL}(x,y) = { \sum_{i}^N {\rm sky}_i (x,y) /\chi_i^2(x,y) \over  {\sum_{i}^N 1/\chi_i^2(x,y)}},
\end{equation}
where ${\rm sky}_i$ and $\chi_i^2$ are the \galfit\ sky background and $\chi^{2}$ values for the $i$-th postage stamp, respectively, and $N\lesssim10$ (Figure~\ref{fig:method}, {\it bottom left}).
Typical fractions of non-source pixel in the stamp (i.e. those pixels are used for estimating the sky) are high ($\sim80\%$), so any background overestimates caused by galaxy light should be small.

We repeat this process for all HFF images, yielding 7\,ICL maps for each cluster from which we derive radial light profiles and SEDs (Sections~\ref{ssec:rad}, \ref{ssec:sed}).

One caveat in our method is that the ICL measurement could be affected if the intrinsic galaxy profile significantly deviates from S\'ersic model.
While the modeled sky in \galfit\ is not directly affected by the residual of the fit, adopting inappropriate models for galaxy profile would lead to wrong estimates of the sky.
Our weighting scheme by the fit goodness ($\chi_{\rm GALFIT}^2$; Equation~\ref{eq:chi}) mitigates against this.

Figure~\ref{fig:method}, {\it right}, shows two histograms of background counts from one cluster tabulated using non-source pixels as defined by the \sext\ segmentation map.
The original image clearly shows an asymmetric tail to positive values, while ICL-subtracted image \footnote{This also includes unmasked light from galaxies.} is more symmetric about zero. 
Fitting using a symmetric function verifies this impression ($\chi_{\rm fit}^2=4.5$ and 3.0, respectively).
ICL over-subtraction would lead to an excess of negative counts compared to the original image, which is not seen. Hence, our ICL subtraction/reconstruction technique appears robust.

We note that using only source mask (e.g., background estimate by \sext) could allow residual light from galaxy wings to contribute to stamp ``background'' counts, and therefore lead to ICL flux overestimates. 
Using \galfit\ mitigates this effect by fitting for and subtracting galaxy light profiles, as demonstrated in the right panel of Figure~\ref{fig:method}.
While we see an excess in positive pixel in the original mosaic, which is contributed from ICL, undetected galaxies, and the wings of masked-but-detected sources, this is significantly suppressed in the ICL-subtracted one.
Although this implies the possibility that our ICL measurement could include light from undetected galaxies, we conclude the amount of those light is negligible to the total ICL mass (Section~\ref{ssec:und}).

Figure \ref{fig:rgb} shows the reconstructed ICL maps. White ellipses denote $R=300$\,kpc---the radius within which we calculate most quantities---measured from each cluster's BCG (Section~\ref{ssec:bcg}).

The above process (1) is agnostic to the form/distribution of the ICL; (2) covers all image pixels, including those containing galaxies in the original data; and (3) mitigates the contribution from the wings of bright galaxies by modeling galaxy light, not by masking them to a given surface brightness threshold. 
Thus, it has significant advantages over previous approaches.

\begin{figure}
\begin{center}
\includegraphics[width=0.85\linewidth, trim = 0.5cm 0cm 0cm 0cm]{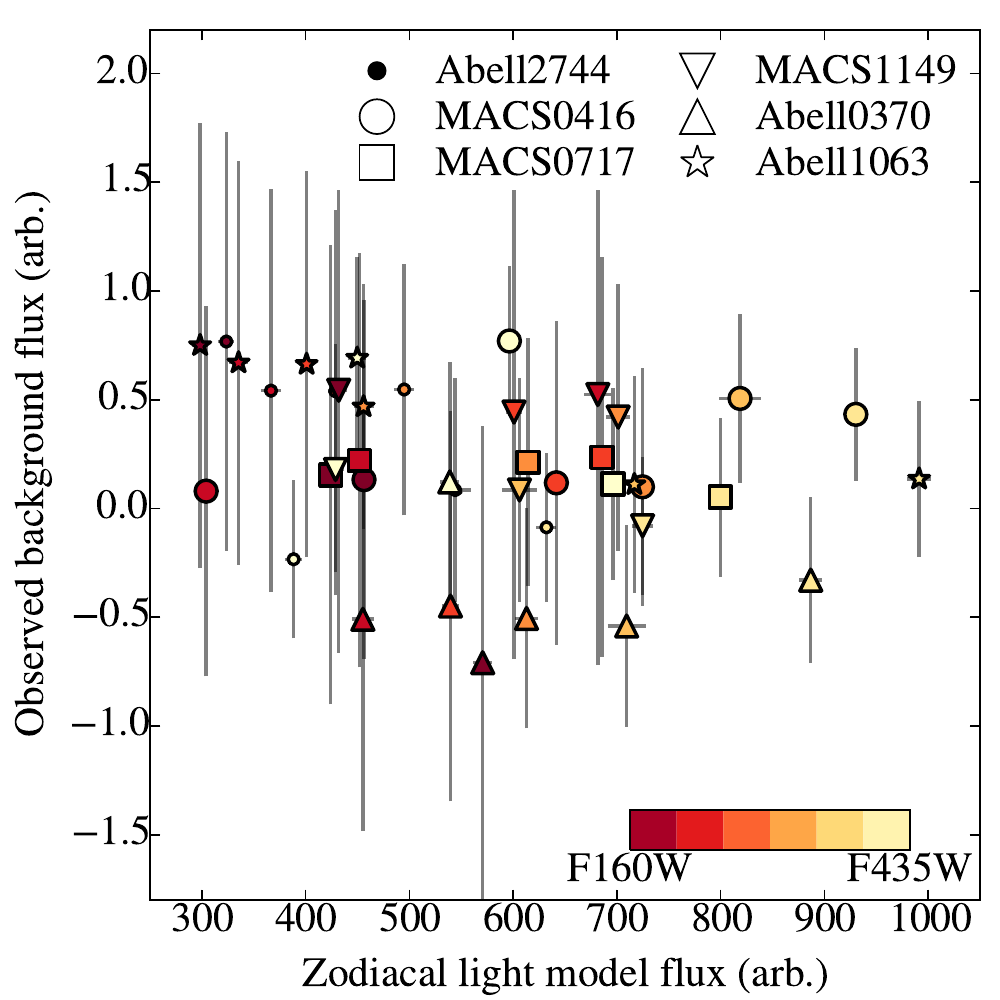}
\caption{
{\bfl  HFF background values ($\mu$; Section~\ref{ssec:bg}, Table~\ref{tab:bg}) plotted against zodiacal light modeled by \citet[][]{kawara17}.
Point colors denote bandpasses (red$\rightarrow$yellow for F160W$\rightarrow$F435W); shapes denote clusters. These values are subtracted from each image, but the absence of significant trends suggests the native HFF pipeline largely removes this potential systematic. Only in Abell0370 do we find signs of oversubtraction.
}
}
\label{fig:zodi}
\end{center}
\end{figure}

\subsection{Uncertainties}\label{ssec:error}

{\bfl Our analysis (Section~\ref{ssec:rad}) ultimately relies on radially averaged ICL properties. Given the large numbers of pixels we are therefore summing over, formal random errors---as inferred from the publicly available HFF RMS maps---are tiny. Additionally, the HFF pipeline accounts for bad pixels and flat-fielding errors. The latter are a $<1\%$ effect \citep{montes14, demaio15} and are also mitigated by the annular averaging. Hence, other uncertainties dominate. 

Principal among these is the global background estimate (Section \ref{ssec:bg}; Table \ref{tab:bg}). 
We discuss all considered sources of error below, but the casual reader may skip to that section. 
Importantly, while this error is systematic on a per-band basis, we are ultimately interested in quantities derived using multiple filters (e.g., stellar mass). Hence, insofar as images from different filters can be treated independently (which we have verified), these actually enter as random noise terms.


\subsubsection{Undetected/Unfitted Light and Sources}\label{ssec:und}
There are two possible contaminants in our ICL measurements --- (1)~faint outer envelope of detected-but-unfitted galaxies and (2)~very faint undetected galaxies.
Masking ensures that detected but unfitted galaxies with $26<m_{160}\lesssim28.5$ (the HFF limiting magnitude; \citealt{lotz17}) do not significantly affect ICL measurements.\footnote{We use the following \sext\ settings: {\sc DETECT\_MINAREA}=9, {\sc DETECT\_THRESH}=1\,$\sigma$, {\sc DEBLEND\_NTHRESH}=64\,$\sigma$, and {\sc DEBLEND\_MINCONT}=0.00001.} 
The total amount of light from {\it undetected} sources---e.g., ultra diffuse galaxies (UDGs)---is also negligible.
For example, Abell2744 and AbellS1063 contain $\sim$50 UDGs in the HFF FoV \citep{lee17}.
The total stellar mass of these objects amounts to $\sim$$10^{10}\, \Msun$, $\simlt10\%$ of the ICL's. Local studies have yielded similar estimates \citep[e.g.,][]{vandokkum15a, koda15b, yagi16}.


\subsubsection{Terrestrial \ion{He}{2} Emission}\label{ssec:he2}

Atmospheric \ion{He}{2} $\lambda1.083\,\mu$m emission is detectable on images taken from earth's day side and affects some F105W data \citep{brammer14, brammer16}. The HFF pipeline removes the average level of this background from each individual exposure, however small-scale spatial variations can remain. Rather than attempt to precisely subtract this light, we simply repeat the analysis in Section~\ref{sec:est} without reference to the F105W data. No significant changes emerge, so we include this imaging below.


\subsubsection{Local Background PSF Sensitivity}\label{ssec:psf}

We rely on \galfit\ galaxy models to provide the local background estimates that ultimately become our ICL maps. \galfit\ in turn relies on knowledge of the instrumental PSF to produce these estimates. We adopt a single image of a bright star in the relevant field as our PSF model. Thus, depending on the background estimate's sensitivity to it, any positional variations in the PSF might lead to large-scale trends in background values and bias ICL maps.

The size of this effect can be tested by burying mock galaxies with typical structural properties (M17) in the image, varying their positions and the PSF stars used in the modeling process, and rederiving local background values as in the initial ICL reconstruction process. 
While we add the mean variation (one per band per cluster) in quadrature to each ICL radial bin's raw photometric uncertainty (derived from the HFF RMS maps), we note that it reaches just $\sim$2\%--20\% of the background uncertainty and is thus dwarfed by that effect.


\subsubsection{Background Subtraction}\label{ssec:bg}

The HFF pipeline provides nominally globally background-subtracted images, removing, e.g., short-timescale fluctuations within each NIR exposure,\footnote{\url{https://blogs.stsci.edu/hstff/2014/06/26/variable-background-signal-in-the-ir-channel-observations/}} and mean terrestrial \ion{He}{2} emission (Section \ref{ssec:he2}). 
Yet, given the presence of the diffuse ICL itself and other sources (diffuse galactic or zodiacal light; \citealt{tsumura13,kawara17}), a concern remains that the HFF pipeline may over- or underestimate the true global pedestal of the mosaic. 

To proceed, we assume that periphery of each image is free from any ICL. We examine all non-source/unmasked pixels outside an ellipse with $R=1200$ pix ($\sim$300\,kpc, or $\sim 0.2\,r_{500}$ for the present sample; see Table~\ref{tab1}) as measured along the BCG's semi-major axis (Section~\ref{ssec:rad}).\footnote{In MACS0416, we manually mask the southern region, where ICL light is prominent.} 
We then fit the resulting pixel distribution with a gaussian, whose mean ($\mu$) and standard deviation ($\sigma$) are summarized in Table~\ref{tab:bg}.
We subtract $\mu$ from all subsequent photometry (Section \ref{ssec:rad}) and fold $\sigma$ into the ICL surface brightness error bar.

It is noted that subtracting the sky measured in the way above could lead underestimation the total amount of light, if the assumption was not correct.
While this is reasonable from previous studies of local clusters \citep[, which are more abundant in ICL; e.g.,][]{mihos17}, a wider FoV observation would be necessary to confirm the validity.

Comparing $\mu$ to the expected backgrounds caused by zodiacal and faint galactic light \citep{kawara17} shows no significant correlation (Figure~\ref{fig:zodi}), implying the initial HFF pipeline largely removed these effects. However, we do find a significant over-subtraction in the original HFF mosaic of Abell0370.
This is understandable: this cluster's apparent size is the largest among the 6 HFF systems, such that the contribution from bright objects and the ICL could affect the HFF pipeline's background estimate.

As to $\sigma$, this reaches $\simlt20\%$ of the ICL's flux at $R\sim200$\,kpc ($\mu_{\rm ICL}\sim27$\,mag\,arcsec$^{-2}$) for the WFC3IR bands and ACS F814/606W. F435W, however, is largely unconstraining outside of cluster central regions.

\begin{figure*}[t!]
\centering
\includegraphics[width=0.8\linewidth]{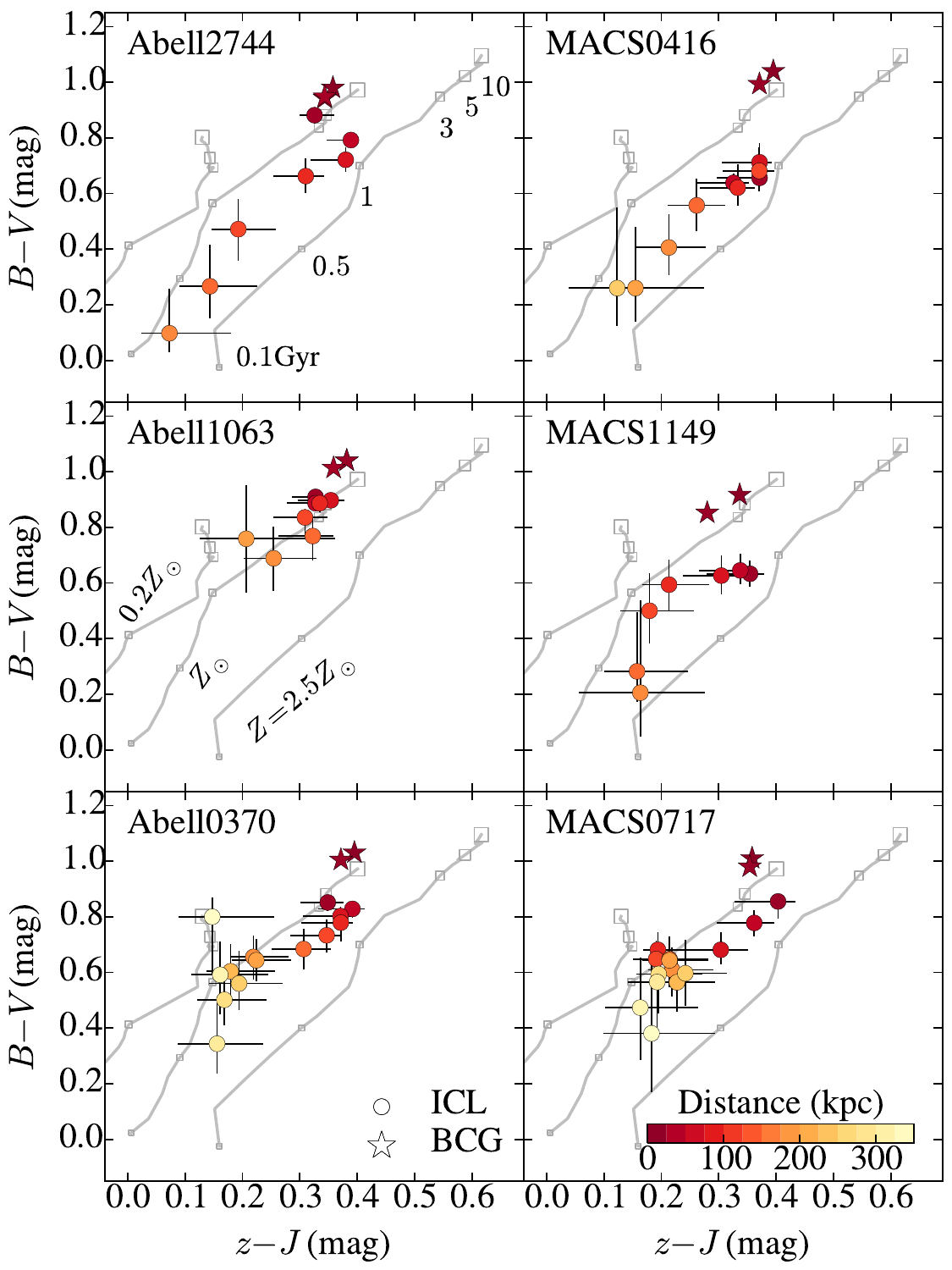}
\caption{
Rest-frame $B-V$ and $z-J$ colors of ICL (circles) and BCGs (stars) in 6\,HFF clusters, binned in $\Delta R=5$\,kpc for BCGs and 25\,kpc for ICL.
The color of the symbols represents the radial distance from the center.
Error bars represent the median error in each bin, scaled by $1/\sqrt{N_{\rm bin}}$, where $N_{\rm bin}$ is the number of pixel in each bin.
The evolution tracks (solid lines) with $Z=2.5\,Z_\odot$ (rightmost), 1\,$Z_\odot$ (center), and $0.2\,Z_\odot$ (leftmost) calculated by \galaxev\ (Chabrier IMF; $\tau\sim0$) are shown.
Ages of each track are tagged with squares (0.1, 0.5, 1, 3, 5, 10\,Gyr from small to large).
}
\label{fig:bvj_rad}
\end{figure*}
\section{Physical Parameter Estimates}\label{sec:est}

\subsection{Selecting Brightest Cluster Galaxies}\label{ssec:bcg}

We set $R=0$ at the location of each cluster's BCG. 
Simply, we define BCGs as the brightest galaxies in F160W from the M17 catalog, with a consistent redshift with each cluster, $|z_{\rm cls}-z_{\rm BCG}|/(1+z_{\rm cls})<3\delta_{z, lim.}$ (where $\delta_{z, lim.}=0.021$; M17).
Table~\ref{tab:bcg} summarizes their properties.
BCG locations roughly agree with the peak of mass density from lens model estimates for all the clusters,\footnote{\url{https://archive.stsci.edu/prepds/frontier/lensmodels/}} and with the X-ray peaks of the relaxed ones (MACS1149 and AbellS1063).

BCG positions visually coincide with ICL peaks except in MACS0717, where the ICL map shows a second peak.
A galaxy of comparable brightness to the BCG resides in this region, but well in front of the cluster ($z_{\rm phot}=0.15\pm{0.02}$) and on the opposite side of the BCG (southeast) relative to the X-ray peak (northwest). 
While it thus does not affect BCG identification, it may bias ICL estimates in that region due to its size ($r>150$\,pixel). Hence, we mask it using a $r=15$\,arcsec circle, excluding those pixels from further analysis.
We hardly see any feature of the peak in the following analyses after masking the region (for example, in Figure~\ref{fig:lm_rad}), and conclude that the foreground galaxy does not significantly affect our results.

We use \fast\ \citep{kriek09} to estimate BCG stellar masses as we do for the ICL and other galaxies (Section~\ref{ssec:sed}).
Because BCG stellar envelopes can extend to $R>100$~kpc, masses derived from \sext's $m_{160}$ auto fluxes (see M17) are often underestimates (by $\sim1$\,mag for our sample; see Table~\ref{tab:bcg}).
We correct for this using the \galfit\ S\'ersic model magnitude ($m_{\rm GALFIT}$), given that the model reproduces the observed BCG profiles \citep[e.g.,][]{zibetti05, ferrarese06}:
\begin{equation}\label{eq:mcor}
	M_*^{\rm cor} = M_*^{\rm SE} \times 10^{(m_{\rm AUTO}-m_{\rm GALFIT})/2.5},
\end{equation}
where $M_*^{\rm SE}$ is the estimate based on \sext\ fluxes. 

\subsection{ICL Radial Profiles}\label{ssec:rad}

We base all physical inferences on azimuthally averaged ICL radial profiles.
These are derived independently in each band before SED-fitting (Section \ref{ssec:sed}).

We adopt the BCG's position as the center of the ICL in each cluster. The radial distance, $R$, is then calculated along the major axis, with the BCG's axis ratio ($q$) and position angle (PA) defining a series of 10 pix wide elliptical apertures for photometry. The ICL flux at each radius is taken as the RMS-weighted average value of the ICL maps in each aperture minus the mean background estimate, $\mu$, in the relevant band (Table \ref{tab:bg}). RMS uncertainties are the quadrature sum of PSF-induced local background fluctuations and the HFF-supplied noise map (Section~\ref{sec:method}).
BCG profiles are derived identically but using 5 pix bins.

Global background subtraction effects are incorporated by perturbing each radial bin by a gaussian random number with a standard deviation of $\sigma$ (Table \ref{tab:bg}). We do this $N=500$ times, leading to 500 ICL profiles per band ready for SED fitting.

\subsection{Radially Resolved SEDs}\label{ssec:sed}

BCG and ICL stellar masses and rest-frame colors (Johnson-Cousins $B$, Bessel $V$, SDSS $z$, and WIRCam/CFHT $J$ bands) are derived using \fast.
For consistency with M17 galaxy properties, we adopt \galaxev\ SSP templates \citep[][hereafter BC03]{bruzual03}, assuming a Chabrier IMF. We assume no dust \citep[e.g.,][]{kitayama09}, and allow metallicities to span $Z/\,Z_\odot\in\{0.2,0.4,1,2.5\}$, where $Z_\odot=0.02$ is solar metallicity.

We fit each of the \nsys\ sets of 7-band radial profiles described in Section \ref{ssec:rad} and adopt the 50th percentile mass and rest-frame color outputs as our best estimates. Uncertainties are defined by the 16th and 84th percentile outputs. These spreads are much larger than those from the formal errors, so we hereafter calculate all derived uncertainties based on them.

\renewcommand{\arraystretch}{1.3}
\tabletypesize{\scriptsize} \tabcolsep=0.03cm
\begin{deluxetable*}{lcccccccccccccccc} \tablecolumns{17}
    \tablecaption{Cluster and Stellar Mass Properties in HFF Clusters.}
    \tablehead{\colhead{Cluster} & \colhead{Redshift} & \colhead{$r_{\rm FoV}$ \tablenotemark{\bfb a}} & \colhead{$r_{500}$ \tablenotemark{\bfb b}} & \colhead{$M_{500}$ \tablenotemark{\bfb b}} & \colhead{$M_{\rm *,300}^{\rm ICL}$ \tablenotemark{\bfb c}} & \colhead{$M_{*,300}^{\rm gal.}$ \tablenotemark{\bfb d}} & \colhead{$M_{*,500}^{\rm gal.}$ \tablenotemark{\bfb d}} & \colhead{$f_{*,300}^{\rm ICL}$ \tablenotemark{\bfb e}} & \colhead{$f_{*,500}^{\rm ICL}$ \tablenotemark{\bfb e}} & \colhead{$f_{*,300}^{\rm BCG}$ \tablenotemark{\bfb e}} & \colhead{$f_{*,500}^{\rm BCG}$ \tablenotemark{\bfb e}} & \colhead{$f_{*,300}^{\rm ICL+BCG}$ \tablenotemark{\bfb e}} & \colhead{$f_{*,500}^{\rm ICL+BCG}$ \tablenotemark{\bfb e}}&\colhead{$B-V$ \tablenotemark{\bfb f}} & \colhead{$z-J$ \tablenotemark{\bfb f}}\\
    \colhead{} & \colhead{} & \colhead{}  & \colhead{} & \colhead{} & \colhead{} & \colhead{$(<300\,{\rm kpc})$}  & \colhead{$(<500\,{\rm kpc})$} & \colhead{$(<300\,{\rm kpc})$} & \colhead{$(<500\,{\rm kpc})$} & \colhead{$(<300\,{\rm kpc})$} & \colhead{$(<500\,{\rm kpc})$} & \colhead{$(<300\,{\rm kpc})$} & \colhead{$(<500\,{\rm kpc})$} & \colhead{$$} & \colhead{$$}\\
    \colhead{} & \colhead{} & \colhead{(Mpc)} & \colhead{(Mpc)} & \colhead{$(10^{14}M_\odot)$} & \colhead{$(10^{11}M_\odot)$}  & \colhead{$(10^{11}M_\odot)$} & \colhead{$(10^{11}M_\odot)$} & \colhead{} & \colhead{} & \colhead{} & \colhead{} & \colhead{} & \colhead{} & \colhead{(mag)} & \colhead{(mag)}
    }
\startdata
{\sc Abell2744} & 0.308 & 0.28 & $1.65\pm{0.07}$ & $17.6\pm{2.3}$ & $1.73_{-0.96}^{+0.95}$	& $13.80_{-3.79}^{+3.33}$	& $15.79_{-3.32}^{+3.55}$	& $0.09_{-0.01}^{+0.02}$	$_{-0.05}^{+0.15}$	& $0.08_{-0.01}^{+0.02}$	$_{-0.04}^{+0.13}$	& $0.17_{-0.03}^{+0.04}$	$_{-0.20}^{+0.18}$	& $0.15_{-0.02}^{+0.03}$	$_{-0.18}^{+0.16}$	& $0.26_{-0.04}^{+0.07}$	$_{-0.25}^{+0.33}$	& $0.24_{-0.03}^{+0.05}$	$_{-0.22}^{+0.30}$	& $0.77_{-0.15}^{0.08}$ & $0.32_{-0.10}^{+0.07}$\\
{\sc AbellS1063} & 0.348 & 0.31 & $1.76\pm0.09$ & $22.5\pm{3.3}$ & $5.69_{-1.57}^{+1.64}$	& $9.76_{-2.01}^{+4.03}$	& $15.51_{-3.37}^{+5.35}$	& $0.23_{-0.03}^{+0.02}$	$_{-0.19}^{+0.29}$	& $0.19_{-0.03}^{+0.02}$	$_{-0.16}^{+0.25}$	& $0.36_{-0.05}^{+0.03}$	$_{-0.42}^{+0.36}$	& $0.29_{-0.04}^{+0.04}$	$_{-0.35}^{+0.31}$	& $0.59_{-0.09}^{+0.05}$	$_{-0.61}^{+0.66}$	& $0.48_{-0.07}^{+0.06}$	$_{-0.51}^{+0.56}$	& $0.87_{-0.13}^{0.05}$ & $0.29_{-0.08}^{+0.08}$\\
{\sc Abell0370} & 0.375 & 0.32 & $1.40\pm0.08$ & $11.7\pm{2.1}$ & $5.37_{-1.10}^{+1.09}$	& $22.71_{-3.99}^{+8.03}$	& $32.87_{-5.81}^{+9.09}$	& $0.15_{-0.03}^{+0.02}$	$_{-0.13}^{+0.18}$	& $0.12_{-0.02}^{+0.02}$	$_{-0.10}^{+0.15}$	& $0.20_{-0.04}^{+0.03}$	$_{-0.21}^{+0.20}$	& $0.16_{-0.03}^{+0.02}$	$_{-0.17}^{+0.16}$	& $0.35_{-0.07}^{+0.05}$	$_{-0.34}^{+0.38}$	& $0.27_{-0.05}^{+0.04}$	$_{-0.27}^{+0.31}$	& $0.74_{-0.17}^{0.10}$ & $0.28_{-0.09}^{+0.09}$\\
{\sc Macs0416} & 0.396 & 0.33 & $1.27\pm0.15$ & $9.1\pm{2.0}$ & $2.37_{-0.69}^{+0.68}$	& $18.58_{-5.98}^{+8.42}$	& $25.04_{-5.11}^{+8.15}$	& $0.09_{-0.02}^{+0.03}$	$_{-0.08}^{+0.13}$	& $0.07_{-0.01}^{+0.01}$	$_{-0.06}^{+0.11}$	& $0.22_{-0.05}^{+0.06}$	$_{-0.27}^{+0.25}$	& $0.18_{-0.03}^{+0.03}$	$_{-0.21}^{+0.20}$	& $0.31_{-0.07}^{+0.09}$	$_{-0.35}^{+0.38}$	& $0.25_{-0.05}^{+0.04}$	$_{-0.27}^{+0.31}$	& $0.66_{-0.13}^{0.13}$ & $0.29_{-0.09}^{+0.09}$\\
{\sc Macs1149}  & 0.544 & 0.40 & $1.53\pm0.08$ & $18.7\pm{3.0}$ & $1.97_{-1.29}^{+1.30}$	& $22.53_{-5.82}^{+6.19}$	& $40.61_{-9.93}^{+14.77}$	& $0.07_{-0.01}^{+0.02}$	$_{-0.03}^{+0.11}$	& $0.04_{-0.01}^{+0.01}$	$_{-0.01}^{+0.06}$	& $0.11_{-0.02}^{+0.03}$	$_{-0.11}^{+0.10}$	& $0.06_{-0.02}^{+0.02}$	$_{-0.06}^{+0.06}$	& $0.17_{-0.03}^{+0.05}$	$_{-0.14}^{+0.22}$	& $0.11_{-0.03}^{+0.03}$	$_{-0.08}^{+0.12}$	& $0.63_{-0.11}^{0.11}$ & $0.26_{-0.08}^{+0.09}$\\
{\sc Macs0717} & 0.548 & 0.40 & $1.69\pm0.06$ & $24.9\pm{2.7}$ & $7.65_{-1.43}^{+1.34}$	& $29.51_{-5.07}^{+12.18}$	& $54.50_{-10.61}^{+12.36}$	& $0.17_{-0.04}^{+0.02}$	$_{-0.15}^{+0.20}$	& $0.11_{-0.02}^{+0.02}$	$_{-0.10}^{+0.13}$	& $0.15_{-0.03}^{+0.02}$	$_{-0.16}^{+0.15}$	& $0.10_{-0.01}^{+0.02}$	$_{-0.10}^{+0.10}$	& $0.33_{-0.07}^{+0.04}$	$_{-0.30}^{+0.35}$	& $0.21_{-0.03}^{+0.04}$	$_{-0.20}^{+0.23}$	& $0.68_{-0.14}^{0.12}$ & $0.26_{-0.09}^{+0.09}$
\enddata
\label{tab1}
\tablenotetext{a}{Physical size of half of the WFC3IR FoV ($60\arcsec$).}
\tablenotetext{b}{\citet{mantz10, sayers13}.}
\tablenotetext{c}{ICL total stellar mass from integrating polynomial fits to $300$\,kpc (Section~\ref{ssec:lmp} and Figure~\ref{fig:lm_rad}). Uncertainties are dominated by estimates derived from F160W light profiles.}
\tablenotetext{d}{Cluster galaxy total stellar mass from integrating polynomial fits to $300$\,kpc or $500$\,kpc. Error is calculated from the fit.}
\tablenotetext{e}{ICL/BCG/ICL+BCG stellar mass over the total cluster stellar mass (ICL+BCG+galaxies) within $R=300$\,kpc and $500$\,kpc. The first error is calculated from the fit and the second is from the uncertainty in the F160W light profiles (error bars and shades of lines in Figure~\ref{fig:frac}, respectively). Due to background subtraction, the ICL mass corresponds to $M_{\rm *,300}^{\rm ICL}$ in all cases (Sections \ref{ssec:bg}, \ref{ssec:lmp}).}
\tablenotetext{f}{Median rest-frame colors (stellar mass-weighted; Figure~\ref{fig:bvzj_mw}). Error represents 16th/84th percentiles.}
\end{deluxetable*}
\renewcommand{\arraystretch}{1.}

\section{Results}

\subsection{ICL Color Gradients}\label{ssec:col}
Figure~\ref{fig:bvj_rad} shows the rest-frame ICL colors of the six HFF clusters.
We use a {\it BVzJ} color-color diagram to examine the age and metallicity of the ICLs' stellar populations. 
This avoids complexities associated with inferring rest-frame $U$-dependent colors due to the poor $S/N$ in the F435W maps.

There is a clear radial ICL color trend such that regions closer to the BCG are redder. 
This gradient suggests older/more mature or more metal-rich stellar populations reside closer to the cluster core while younger ones lie nearer the outskirts, which is consistent with previous findings at the similar redshift range \citep[e.g.,][]{demaio15}. 

Some previous studies have attributed these trends purely to metallicity} and not age.
Because time weakens age but not metallicity gradients by definition, this is a reasonable assumption.
However, at $z\sim0.5$, when ICL formation is ongoing \citep[e.g.,][]{murante07, conroy07, contini14}, it is not obviously correct.

Indeed, though uncertainties are sizable, Figure~\ref{fig:bvj_rad} suggests that the color gradient is not purely a metallicity effect. 
The blue populations at large radii ($R>150$\,kpc) in some clusters (e.g., Abell2744) seem inexplicable by an old-yet-low-$Z$ stellar population. 
This potentially signals an interesting (if unsurprising) change in the ICL makeup between today and epochs nearer to the peak of its formation. 
We return to this point in Section~\ref{sec:discussion}.

To clarify the ICL's relation to the BCG, we show the colors of the latter as stars in Figure \ref{fig:bvj_rad}.
The radial range is up to $\sim$30\,kpc, the limit of the \sext\ segmentation map bounds for these sources.
Most BCGs occupy the {\it upper-right} region of the diagram, implying even more mature stellar populations than those in the ICL, or, conversely, that the BCG and ICL have distinct formation pathways.

\begin{figure*}[t!]
\centering
\includegraphics[width=0.9\linewidth]{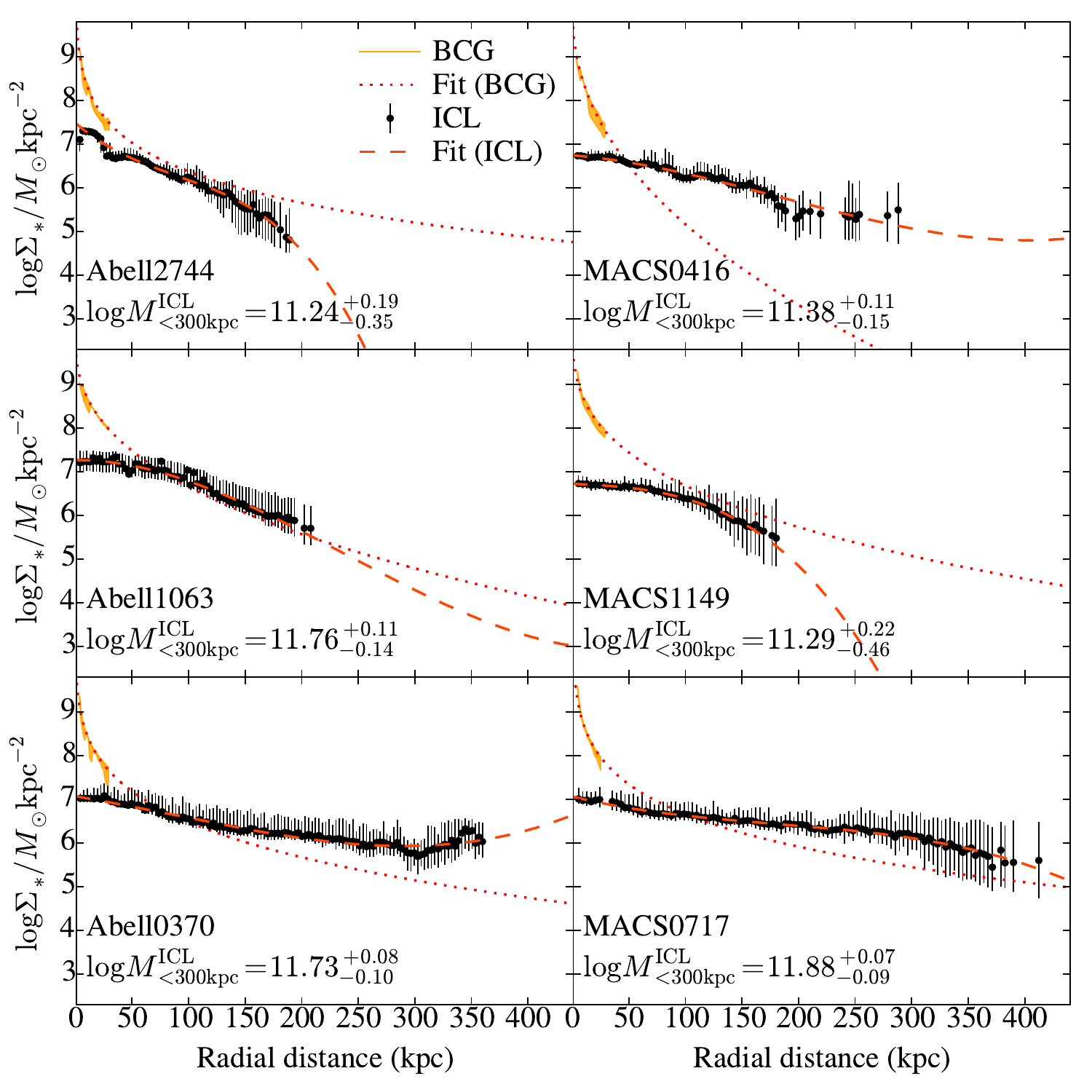}
\caption{
Radial ICL stellar mass profiles.
Red dashes show third-order polynomial fits.
The total stellar mass within $<300$\,kpc, $M_{*, 300}^{\rm ICL}$---calculated by integrating the polynomial fit---is printed in each panel.
BCG radial profiles are shown in yellow, with \galfit\ results as dotted red lines. These deviate from the ICL profiles, suggesting meaningful physical differentiation between these components.
}
\label{fig:lm_rad}
\end{figure*}

\subsection{ICL Stellar Mass Profiles}\label{ssec:lmp}
Figure~\ref{fig:lm_rad} shows the ICL and BCG radial stellar mass surface density ($\Sigma_{*}$) profiles.
The ICL reaches $\lsig\sim7$ in the innermost region, about 100$\times$ lower than that of BCGs ($\lsig\simgt9$). 
The ICL's density then smoothly decreases to $\lsig\sim4$--5 at $R>150$\,kpc. 

To discuss radial trends and calculate total ICL stellar masses, we fit the mass profiles with a 3rd-order polynomial.
This allows us to compare properties among the 6\,clusters (and with previous results), whose limiting surface mass densities vary. Best fit curves are shown in each panel of Figure~\ref{fig:lm_rad}.

We calculate the total ICL stellar mass within $R<300$\,kpc, $M_{*,300}^{\rm ICL}$, by integrating the polynomial fits.
These estimates---which span $\logm \sim 11$ to 12---are printed in each panel and summarized in Table~\ref{tab1}.
Uncertainties are estimated by adding the results from integrating $10^5$ Monte Carlo realizations of the fits (with points modulated by their error bars) in quadrature to the fractional error in the F160W light profile (which should be a good mass proxy). As the formal fitting errors are $\sim$0.03 dex, the latter dominates.

Two clusters (Abell0370 and MACS0717) show markedly shallower radial profiles relative to the others.
This may imply that the ICL in these clusters has been stirred-up by recent mergers \citep{ebeling04, richard10}.

BCG radial profiles are also shown for comparison, with their \galfit\ profiles overplotted.
Note that the large-radius BCG profile extrapolations typically diverge from the ICL, suggesting that we are indeed probing two meaningfully distinct stellar components in these clusters. 
The same holds vis-\`a-vis the cluster {\it galaxy} mass profile, (not plotted).


\begin{figure*}[t!]
\begin{center}
\includegraphics[width=0.8\linewidth]{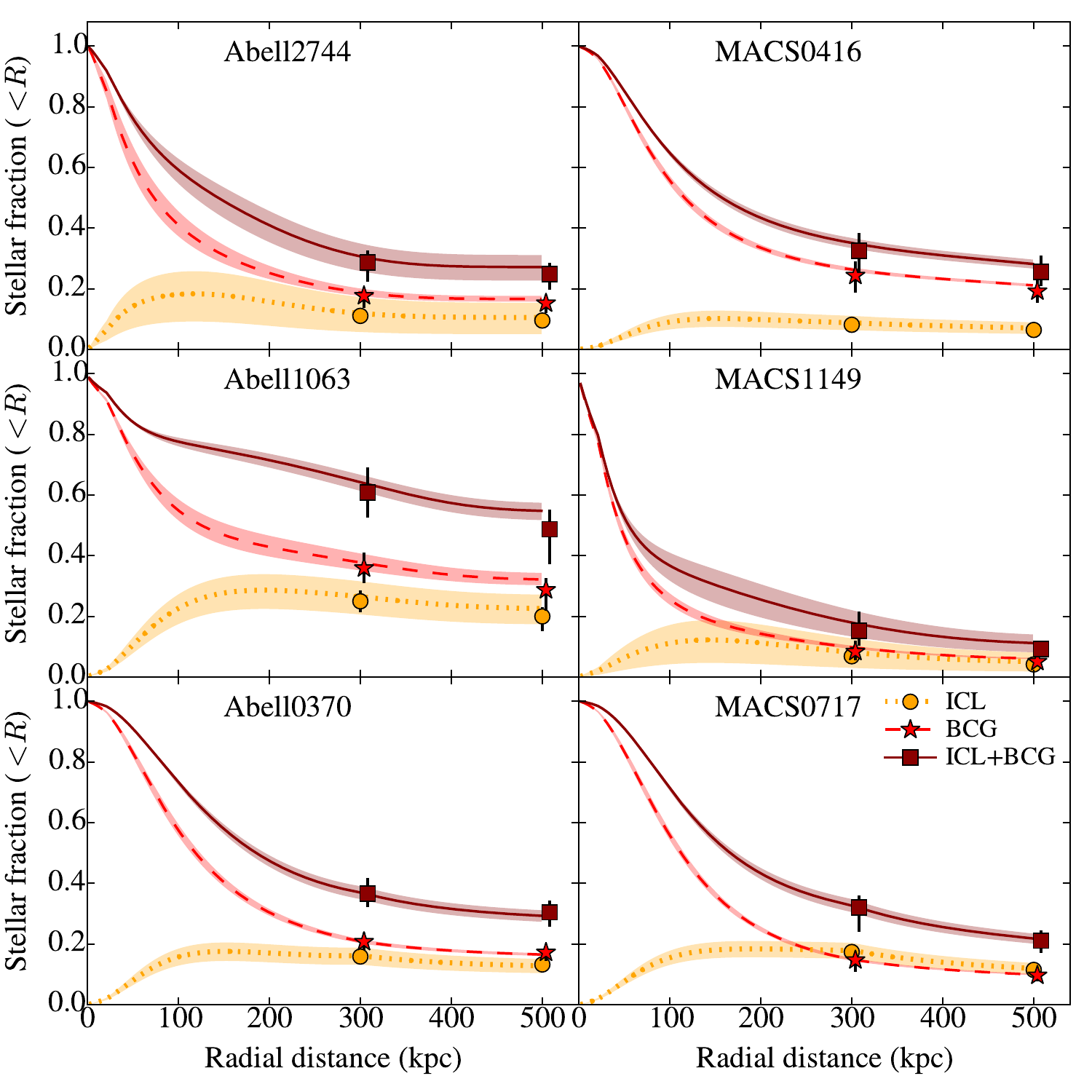}
\caption{
Cumulative fractions of stellar mass in ICL (red lines), BCG (green lines) and ICL+BCG (blue lines) over the total stellar mass (ICL+BCG+cluster galaxies).
The slopes are the median of $10^3$ MCMC realizations, which are not shown here to keep the visual clearness.
The shaded region of the lines represents the uncertainty in the total mass in ICL, while the error bars at $R=300$\,kpc and 500\,kpc do the uncertainties from the polynomial fit (in Table~\ref{tab1}).
It is noted that the region at $R\simgt300$\,kpc is beyond the FoV of \hst\ and our sensitivity to ICL, and we truncate the mass estimate for ICL, while we keep summing the mass in  galaxies by interpolating the polynomial fit.
}
\label{fig:frac}
\end{center}
\end{figure*}


\begin{figure*}[t!]
\centering
\includegraphics[width=0.49\linewidth]{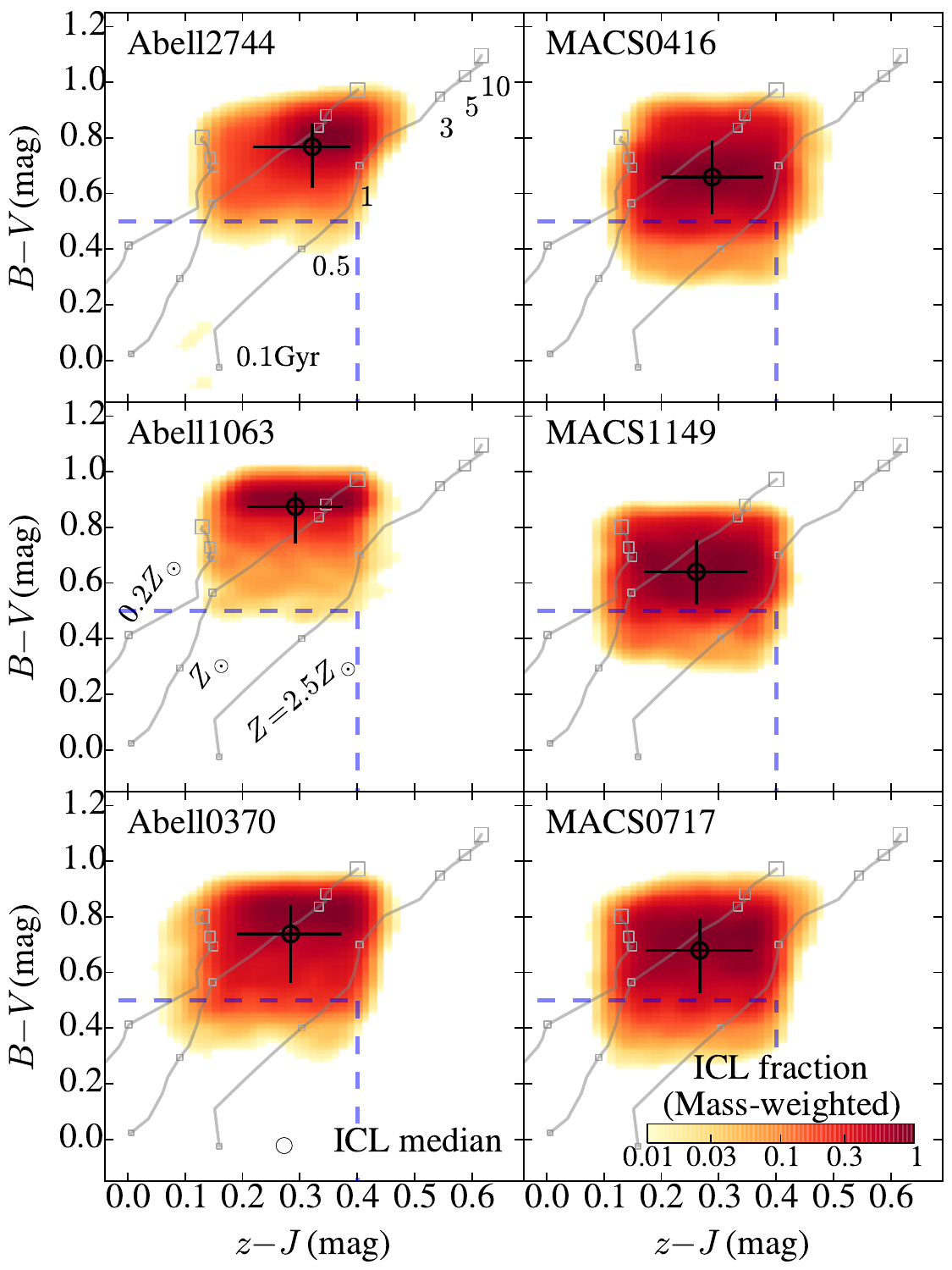}
\includegraphics[width=0.49\linewidth]{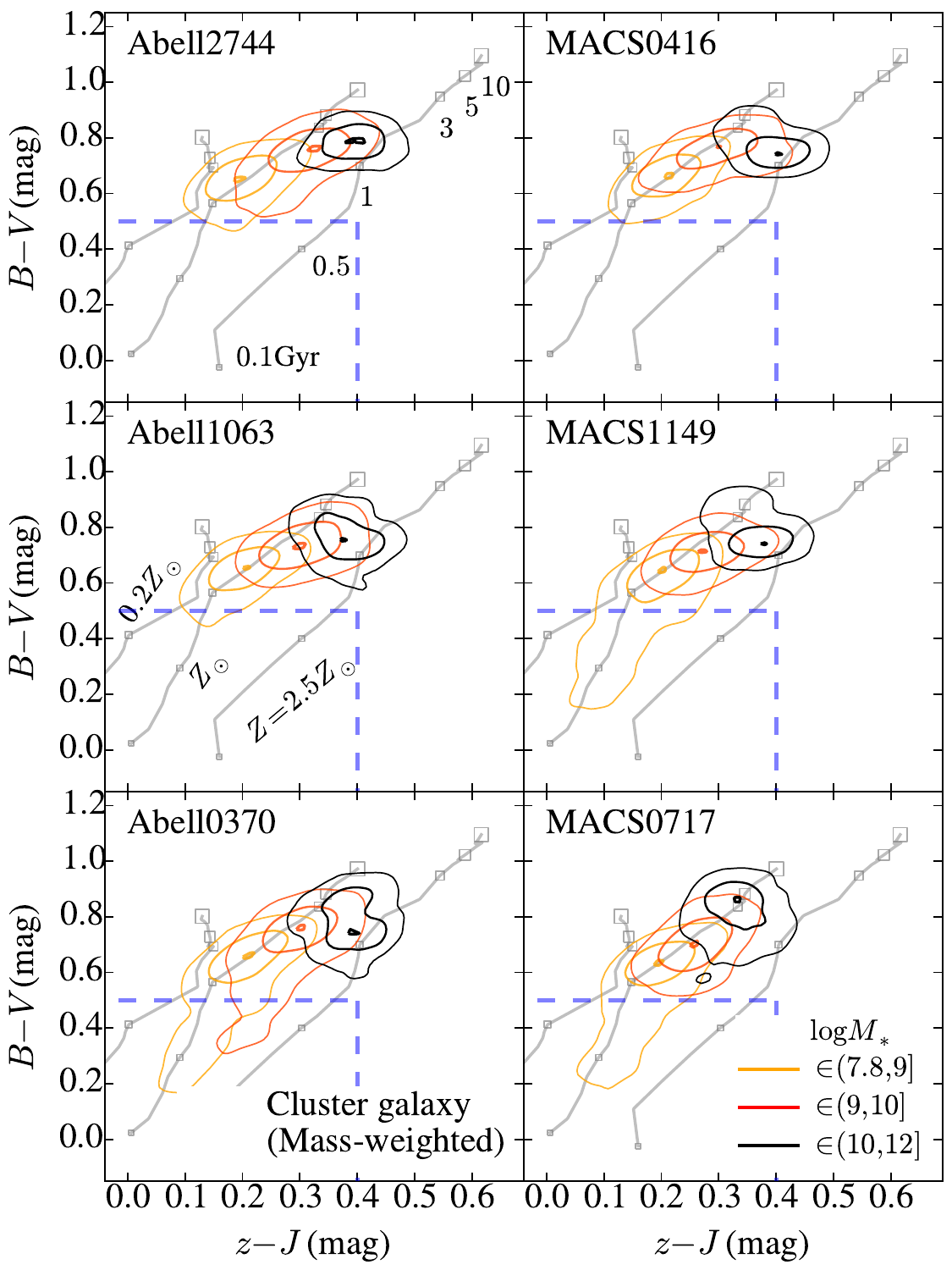}
\caption{
$Left$: Same color-color diagram as Figure~\ref{fig:bvj_rad}, but showing the stellar mass-weighted density map of ICL colors.
The median and its 16/84th percentiles are shown in each panel (circles with error bars), and summarized in Table~\ref{tab1}.
While the ICL color is dominated by old stellar population in most of our clusters, we still see nonnegligible contribution ($\sim5$-10\,\% relative to the old population) from the young population at the left-bottom region (blue dashed lines) in most of our clusters, except for Abell S1063 ($<1\%$).
$Right$: Mass weighted color-color diagram, same as left panel, but for the cluster member galaxies with $\logm\in(7.8,9]$ (orange contours), $(9,10]$ (red), and $(10, 12]$ (black).
Each of three lines of each contour represents the region where 1\,\% (innermost), 50\,\% (middle), 90\% (outermost) of the each population is included.
}
\label{fig:bvzj_mw}
\end{figure*}

\subsection{ICL and BCG Stellar Mass Fractions}\label{ssec:sf}
The ICL mass fraction is an important metric of how ICL formation proceeds with respect to cluster galaxy evolution \citep[][]{conroy07, behroozi13, contini14}. We first calculate the ICL and ICL+BCG mass compared to the total stellar mass in the clusters (ICL+BCG+galaxies). 
We adopt $M_{*}^{\rm cor}$ for the BCG (Equation \ref{eq:mcor}, Table~\ref{tab:bcg}), and $M_{\rm *,300}^{\rm ICL}$ (Table~\ref{tab1}) for the ICL. 
Contributions from the ICL latter are small at $R\simgt300$ kpc (Figure~\ref{fig:lm_rad}), but see below.

The total mass in galaxies is based on the M17 galaxy catalog, complete to $\logm\sim7.8$.
Because the \hst\ FoV ($\lesssim$400 kpc) does not capture the entire extent of the HFF clusters ($R_{500}>1$\,Mpc), the total stellar mass of {\it observed} cluster galaxies will underestimate the total cluster galaxy stellar mass. We correct for this by constructing an azimuthally averaged cluster galaxy mass profile based on the observed systems as we did for the ICL. 
We fit this with a polynomial and integrate it out to $R=500$ kpc, where the fit seems reliable.
The only modification we make from Section \ref{ssec:rad} is to use circular annuli to account for the fact that galaxies can randomly move inside the cluster \citep[e.g.,][]{west17}, rather than elliptical ones based on the BCG's orientation and axis ratio.

The cumulative radial stellar fraction for component $c$ (ICL/BCG/ICL+BCG) is then calculated as:
\begin{equation}
\begin{aligned}
	f_{*,c}(R)=\frac{M_{*, c}(<R)}{M_{\rm *,tot}(<R)},
\end{aligned}
\end{equation}
where the ``tot'' subscript signifies all stellar components (ICL+BCG+galaxies). We derive $f_{*,c}$ for each of the ICL and galaxy profile fits and quote the median as the best estimate.
Figure~\ref{fig:frac} shows the results.

This figure reveals the total ICL+BCG mass fractions to range from $\sim$15\%--60\%, with $\sim$5\%--20\% coming from the ICL alone (to $R\sim300$\,kpc). 
These results are consistent with a previous estimates for one HFF cluster \citep[Abell2744;][]{jimenez16}, but slightly higher than those done with CLASH data \citep[AbellS1063 and MACS0416;][]{burke15}.

For example, the difference in depth between CLASH and HFF \citep[$\Delta m\sim1$\,mag in F160W;][]{postman12,lotz17} can be interpreted that previous studies reach 25\,\% shallower in surface mass profiles.
Missing these faint light (at $\sim150$\,kpc in Figure~\ref{fig:lm_rad}) could lead to $\Delta M_*\sim10^{10}\,M_\odot$ loss in the ICL mass, which corresponds $\sim10$\,\% of our ICL measurement and would explain the discrepancy between ours and previous result from CLASH data.
Fitting and interpolation in the radial mass profile, adopted in this study, could compensate the undetected outer stellar component for those shallower data.
Beside this, the $M/L_{\rm F160W}$ of ICL population varies by a factor of $\sim3$ over the explored radii in this study, which implies that adopting one constant number could also result in under/overestimates in ICL mass.

Regardless, we qualitatively concur with those authors' findings regarding the need for substantial ICL mass growth between $z\sim1$ and the present day (below, and Section \ref{ssec:formation}).

We can say that at least part of the intra-sample variation in mass fractions seems attributable to redshift effects. Examining the two most massive clusters in this study illustrates this: AbellS1063 has the highest ICL+BCG fraction at fixed radius and the second-lowest redshift in the sample. The similarly massive MACS0717 (Table \ref{tab1}) has only about half the ICL+BCG fraction, but the highest redshift in the sample, suggesting the ICL mass may double after $z\sim0.5$ \citep[][]{contini14}.

\section{Discussion}\label{sec:discussion}
\subsection{Interpreting ICL Color Gradients}\label{ssec:cg}

Based on BCG spectra \citep{coccato11} and the colors of individual ICM stars \citep{williams07}, previous low-$z$ work has found the ICL to comprise mainly very old ($\sim10$\,Gyr) G, K, and M stars. If so, the radial color trend in Figure \ref{fig:bvj_rad} would imply a metallicity gradient in the ICL.

However, because these findings come from local clusters, $\sim$5\,Gyr older than those studied here, their applicability in the HFF context is unclear. 
Further, they were limited to very central ICL regions ($\simlt50$\,kpc), where stellar properties may diverge from those at larger radii due to processes associated with BCG formation.

Indeed, we find blue stellar populations in the outer ICL which do not seem explainable by metallicity effects in an ancient stellar population. 
Rather, these suggest the presence of a significant number of stars with ages $\sim1$\,Gyr, as previously reported from imaging, and also spectroscopy \citep[e.g.,][]{edwards16}. We therefore posit that these regions comprise sizable numbers of stars of A-type or earlier stripped from starforming galaxies on cluster infall \citep[][]{abramson11, poggianti16,vulcani16a,vulcani16b, mcpartland16}---or the perhaps 25\% of $z\sim0.5$ cluster galaxies that are poststarbursts/H$\delta$ strong starbursts with similar stellar ages \citep[e.g.,][and references therein]{poggianti09,Dressler13}.
As $z\rightarrow0$, this blue light would diminish as the population passively evolves over the intervening $\sim$5 Gyr \citep[e.g.,][]{coccato10, mihos17}.\footnote{Modulo small additions via, e.g., induced star formation in ram pressure-stripped gas \citep[e.g.,][]{yoshida02, yagi13}.}

To better understand the importance of these stars to the global cluster ICL, we modify the color-color diagram of Figure~\ref{fig:bvj_rad} by weighting the colors by the total stellar mass in their radial bin.
Figure~\ref{fig:bvzj_mw} shows the updated diagram.
Here, we see that, while the ICL is dominated (by mass) by slightly redder/older stars (ages $\sim$1--3 Gyr), the younger component still contributes meaningfully in most of the clusters.
We find the mass contained in this population (defined by $B-V<0.5$ and $z-J<0.4$) is $\sim5$--10\,\% of the ICL's total mass, except for one cluster (Abell S1063; $<1\,\%$).
That is, $\logm\sim10.4$--11.2 of stellar mass is locked in these young stellar populations, especially at the large radii, qualitatively consistent with the galaxy distribution in clusters at this redshift range \citep[e.g.,][]{dressler97, koyama11}.
As the number of star forming galaxies in clusters decreases with time \citep{butcher78, kodama01}, the supply of the young stellar population will itself also decrease. 
We discuss this further in Section \ref{ssec:formation}. 
Note that elliptical galaxies exhibit a similar phenomenon \citep[e.g.,][]{treu05} wherein old stellar populations have a $\sim$10\% (by mass) frosting of young stars at intermediate redshifts.

As for Abell S1063, the lack of such a young population is understandable: this is the most relaxed cluster in the HFF sample \citep{lotz17}, such that dynamical events---and hence maybe the associated and production of new stars (see below)---should be less frequent.

\subsection{The Formation of the ICL}
\label{ssec:formation}
Comparing our measurements to local values \citep[20-40\%;][though in less-massive clusters]{gonzalez13} and numerical simulations \citep{martel12, contini14}, ICL mass fractions should roughly double between $z\sim0.5$ and today. 
This would entail the addition of $\Delta M_{*}\sim2\times10^{11}\, \Msun$ to the ICL from some source\footnote{We assume that the observed ICL evolution is attributed to the redshift evolution, rather than biases in comparison. This is reasonable given the fact that the present sample and the local sample are both the most massive clusters at each epoch.}. 
Simply dividing that amount by the time between the two epochs, the implied growth rate is $dM_*/dt\sim2\times10^{11}\,\Msun/5\, {\rm Gyr}=40\,\Msun\,{\rm yr^{-1}}$. 
This is a relatively high {\it absolute} rate, but since clusters harbor hundreds to thousands of galaxies, it can be sustained by relatively small leaks on a per-galaxy basis.

The stars giving rise to this mass growth of the ICL potentially come from (1) newly accreted galaxies, and/or (2) the ``post-processing'' of already acquired, likely passive, cluster galaxies \citep[e.g.,][]{zabludoff98, birrier09, vija13}.

Under the first scenario, since many of the newly accreted galaxies would have been recently starforming \citep[e.g.,][]{Dressler13}, we might expect the ICL colors to {\it remain} relatively blue compared to the mature cluster galaxy population, as it was observed to be {\it at} $z\sim0.5$ (Section \ref{ssec:cg}).

On the other hand, if the growth is mainly due to post-processing, there would be no substantive blue population from which to draw ICL stars. As such, passive evolution of the older ICL component and the source galaxies should redden the observed ICL over time.

Studies of the Virgo cluster---which is less massive and therefore, if anything, younger than the descendants of our clusters---show its ICL to have $B-V\sim0.8$-1 \citep[][]{castro-rodrigu09, rudick10, mihos17}.
Other studies beyond the local universe but at lower redshifts \citep[$z\lesssim0.3$;][]{zibetti05, toledo11, presotto14} also show redder colors than most of what we observe.
Our SSP calculations show passive evolution of the observed population would be consistent with the Virgo and other low-$z$ clusters' ICL color \citep[e.g.,][]{krick07} by $z\sim0$ (Figures \ref{fig:bvj_rad}, \ref{fig:bvzj_mw}).
Hence, it would appear that, at least at {\it later} times ($z\lesssim0.5$), ICL growth is driven substantially by the removal of stars from the established cluster population, rather than newly infalling, starforming systems.

The origin of the {\it observed} ICL in the HFF sample is hazier. 
The presence of young stellar populations (ages $\sim$1 Gyr) in some of our clusters suggests significant recent growth in the ICL. This seems inescapably tied to the presence of recently starforming or recently acquired galaxies (including [post-]starbursts). But, the $z\sim0.5$ ICL's mass is dominated by stars with older ages---$\sim 1$-3\,Gyr (Figure~\ref{fig:bvzj_mw}). This baseline population implies ICL construction began at $z\lesssim1$ (consistent with a numerical calculation by \citealt{contini14}), but from what was it built?

The {\it BVzJ} diagram would suggest lower mass {\it passive} ($\logm\simlt10$) cluster galaxies may be an important source. The ICL is more consistent with those systems' colors than it is with those of (at least the cores of) more massive galaxies or BCGs, which are $\sim$0.3 mag redder in $B-V$ and therefore presumably much older. 
Given that lower-mass cluster galaxies are also much more abundant than their higher-mass counterparts (e.g., M17), this finding would argue for the ICL to come more from the stripping/cannibalism \citep[e.g.,][]{ostriker75, moore96, treu03, nipoti03, nipoti04} of low-mass ($\logm\lesssim10$), as opposed to even Milky Way mass galaxies.\footnote{Another interesting comparison is to the colors of UDGs, which are slightly bluer than other bright cluster galaxies \citep[$\Delta {\rm F814W-F105W}\sim0.2$;][]{lee17}, and consistent with our ICL color measurements (Figure~\ref{fig:bvzj_mw}).}
Perhaps half of these were already in the cluster at $z\sim1$ (M17; see their Figures 9, 11), with the remainder having been accreted subsequently from both passive and starforming populations (Section~\ref{ssec:cg}), and so could indeed serve as a substantial ICL source population.

This interpretation is partly in tension with that of \citet[][]{montes14} and predictions of \citet{contini14}, which suggest that Milky Way mass ($\logm\gtrsim10.5$) galaxies are the most likely ICL source based on dynamical friction arguments, in terms of the observed colors. 
While the integrated color of these massive galaxies are typically redder than the observed ICL (Figure~\ref{fig:bvzj_mw}), the presence of strong color gradients in such galaxies could remedy this discrepancy since the ICL (especially at larger clusto-centric radii) is likely to come from easily stripped stars at large galactocentric radii, which would also be the bluest \citep[e.g., ][Figure 9]{morishita15}.
Additionally, the ICL mass in our clusters is roughy equal to that of all {\it extant} $\logm<10$ red galaxies (M17), implying that some help from higher-mass systems is likely, as favored by \citet{montes14}. Hence, dynamical friction acting on more massive galaxies might source the ICL nearer to BCGs, while lower mass, bluer galaxies source the rest of it.

This being said, it seems secure to posit from both their colors (Figure \ref{fig:bvj_rad}) and mass profiles (Figure \ref{fig:lm_rad}) that the ICL does not arise through the same mechanisms that establish at least the BCG core, and likely  most of the high-mass ``native'' cluster galaxy population.

Hence, the ICL appears to have arisen at relatively late times ($z\lesssim1$). 
We suggest its early growth was driven by a mix of the stripping of infalling galaxies and the extant passive cluster population, with the mix shifting mainly towards the latter process at later times ($z\lesssim0.5$; given the red ICL colors of $z\sim0$ clusters).
Deep spatially resolved spectroscopy would help disentangle age/metallicity degeneracies and thus perhaps clarify this picture.


\section{Summary}\label{sec:summary}

\begin{enumerate}
	\item Based on a new method and very deep multi-band HFF imaging, we derive the radial light profiles of 6 HFF clusters at $0.31<z<0.55$. 
	\item Via SED fitting, we calculated the stellar mass and rest-frame colors of the ICL and BCGs in these systems out to $R\sim300$ kpc.
	\item We find the ICL's mass to be dominated by moderately old stellar populations ($\sim$1--3 Gyr). From their colors, these are consistent with having been drawn from stripped quiescent cluster galaxies of $\logm\simlt9.5$ starting at $z\sim1$.
	\item However, prominent color gradients in all systems reveal stellar populations at $R\simlt150$\,kpc  that are too blue to be explained without invoking $\sim$5\%--10\% mass fractions of A and earlier-type stars (age $\sim1$ Gyr), presumably drawn from recently starforming/infalling galaxies.
	\item The stellar mass in the ICL is $\sim10\%$ of the total stellar components in clusters (ICL+BCG+cluster galaxies; $R\lesssim500$ kpc), about a half of the local value, suggesting that the ICL is still under construction at the redshifts observed.
	\item Since then, given the redder colors seen in local clusters, perhaps $\sim40\, \Msun\, {\rm yr^{-1}}$ must be dumped into the ICL, likely drawn from the extant passive cluster population.
\end{enumerate}

These findings emphasize the importance of the cluster-specific mechanisms on galaxies at the early epoch of the ICL formation, which might lead to the existence of starburst galaxies in the cluster environment at this redshift, and on low-mass passive galaxies, in addition to the merging of more massive galaxies, at the later epoch after the peak of the cosmic star formation activity.

\acknowledgements
We thank the anonymous referee for constructive comments.
We also acknowledge Dr. K.\ Tsumura (Tohoku university) for providing the model of zodiacal light, and Drs.\ C.\ Y.\ Peng (GMTO) and T.\ Ichikawa (Tohoku university) for insightful advices.
Support for this work is provided by NASA through HST-GO-13459. 
T.M. acknowledges support from the Japan Society for the Promotion of Science (JSPS) through JSPS research fellowships for Young Scientists.
B.V. acknowledges the support from an Australian Research Council Discovery Early Career Researcher Award (PD0028506).

\bibliographystyle{apj}
\bibliography{./adssample}

\begin{thebibliography}{}
\expandafter\ifx\csname natexlab\endcsname\relax\def\natexlab#1{#1}\fi

\bibitem[{{Abramson} {et~al.}(2011){Abramson}, {Kenney}, {Crowl}, {Chung}, {van
  Gorkom}, {Vollmer}, \& {Schiminovich}}]{abramson11}
{Abramson}, A., {Kenney}, J.~D.~P., {Crowl}, H.~H., {et~al.} 2011, \aj, 141,
  164

\bibitem[{{Behroozi} {et~al.}(2013){Behroozi}, {Wechsler}, \&
  {Conroy}}]{behroozi13}
{Behroozi}, P.~S., {Wechsler}, R.~H., \& {Conroy}, C. 2013, \apj, 770, 57

\bibitem[{{Berrier} {et~al.}(2009){Berrier}, {Stewart}, {Bullock}, {Purcell},
  {Barton}, \& {Wechsler}}]{birrier09}
{Berrier}, J.~C., {Stewart}, K.~R., {Bullock}, J.~S., {et~al.} 2009, \apj, 690,
  1292

\bibitem[{{Bertin} \& {Arnouts}(1996)}]{bertin96}
{Bertin}, E., \& {Arnouts}, S. 1996, \aaps, 117, 393

\bibitem[{{Brammer} {et~al.}(2014){Brammer}, {Pirzkal}, {McCullough}, \&
  {MacKenty}}]{brammer14}
{Brammer}, G., {Pirzkal}, N., {McCullough}, P., \& {MacKenty}, J. 2014,
  {Time-varying Excess Earth-glow Backgrounds in the WFC3/IR Channel}, Tech.
  rep.

\bibitem[{{Brammer} {et~al.}(2016){Brammer}, {Marchesini}, {Labb{\'e}},
  {Spitler}, {Lange-Vagle}, {Barker}, {Tanaka}, {Fontana}, {Galametz},
  {Ferr{\'e}-Mateu}, {Kodama}, {Lundgren}, {Martis}, {Muzzin}, {Stefanon},
  {Toft}, {van der Wel}, {Vulcani}, \& {Whitaker}}]{brammer16}
{Brammer}, G.~B., {Marchesini}, D., {Labb{\'e}}, I., {et~al.} 2016, \apjs, 226,
  6

\bibitem[{{Bruzual} \& {Charlot}(2003)}]{bruzual03}
{Bruzual}, G., \& {Charlot}, S. 2003, \mnras, 344, 1000

\bibitem[{{Burke} {et~al.}(2015){Burke}, {Hilton}, \& {Collins}}]{burke15}
{Burke}, C., {Hilton}, M., \& {Collins}, C. 2015, \mnras, 449, 2353

\bibitem[{{Butcher} \& {Oemler}(1978)}]{butcher78}
{Butcher}, H., \& {Oemler}, Jr., A. 1978, \apj, 219, 18

\bibitem[{{Castro-Rodrigu{\'e}z} {et~al.}(2009){Castro-Rodrigu{\'e}z},
  {Arnaboldi}, {Aguerri}, {Gerhard}, {Okamura}, {Yasuda}, \&
  {Freeman}}]{castro-rodrigu09}
{Castro-Rodrigu{\'e}z}, N., {Arnaboldi}, M., {Aguerri}, J.~A.~L., {et~al.}
  2009, \aap, 507, 621

\bibitem[{{Chabrier}(2003)}]{chabrier03}
{Chabrier}, G. 2003, \pasp, 115, 763

\bibitem[{{Coccato} {et~al.}(2010){Coccato}, {Gerhard}, \&
  {Arnaboldi}}]{coccato10}
{Coccato}, L., {Gerhard}, O., \& {Arnaboldi}, M. 2010, \mnras, 407, L26

\bibitem[{{Coccato} {et~al.}(2011){Coccato}, {Gerhard}, {Arnaboldi}, \&
  {Ventimiglia}}]{coccato11}
{Coccato}, L., {Gerhard}, O., {Arnaboldi}, M., \& {Ventimiglia}, G. 2011, \aap,
  533, A138

\bibitem[{{Collins} {et~al.}(2009){Collins}, {Stott}, {Hilton}, {Kay},
  {Stanford}, {Davidson}, {Hosmer}, {Hoyle}, {Liddle}, {Lloyd-Davies}, {Mann},
  {Mehrtens}, {Miller}, {Nichol}, {Romer}, {Sahl{\'e}n}, {Viana}, \&
  {West}}]{collins09}
{Collins}, C.~A., {Stott}, J.~P., {Hilton}, M., {et~al.} 2009, \nat, 458, 603

\bibitem[{{Conroy} {et~al.}(2007){Conroy}, {Wechsler}, \&
  {Kravtsov}}]{conroy07}
{Conroy}, C., {Wechsler}, R.~H., \& {Kravtsov}, A.~V. 2007, \apj, 668, 826

\bibitem[{{Contini} {et~al.}(2014){Contini}, {De Lucia}, {Villalobos}, \&
  {Borgani}}]{contini14}
{Contini}, E., {De Lucia}, G., {Villalobos}, {\'A}., \& {Borgani}, S. 2014,
  \mnras, 437, 3787

\bibitem[{{DeMaio} {et~al.}(2015){DeMaio}, {Gonzalez}, {Zabludoff}, {Zaritsky},
  \& {Brada{\v c}}}]{demaio15}
{DeMaio}, T., {Gonzalez}, A.~H., {Zabludoff}, A., {Zaritsky}, D., \& {Brada{\v
  c}}, M. 2015, \mnras, 448, 1162

\bibitem[{{Dressler} {et~al.}(2013){Dressler}, {Oemler}, {Poggianti},
  {Gladders}, {Abramson}, \& {Vulcani}}]{Dressler13}
{Dressler}, A., {Oemler}, Jr., A., {Poggianti}, B.~M., {et~al.} 2013, \apj,
  770, 62

\bibitem[{{Dressler} {et~al.}(1997){Dressler}, {Oemler}, {Couch}, {Smail},
  {Ellis}, {Barger}, {Butcher}, {Poggianti}, \& {Sharples}}]{dressler97}
{Dressler}, A., {Oemler}, Jr., A., {Couch}, W.~J., {et~al.} 1997, \apj, 490,
  577

\bibitem[{{Ebeling} {et~al.}(2004){Ebeling}, {Barrett}, \&
  {Donovan}}]{ebeling04}
{Ebeling}, H., {Barrett}, E., \& {Donovan}, D. 2004, \apjl, 609, L49

\bibitem[{{Edwards} {et~al.}(2016){Edwards}, {Alpert}, {Trierweiler},
  {Abraham}, \& {Beizer}}]{edwards16}
{Edwards}, L.~O.~V., {Alpert}, H.~S., {Trierweiler}, I.~L., {Abraham}, T., \&
  {Beizer}, V.~G. 2016, \mnras, 461, 230

\bibitem[{{Ferrarese} {et~al.}(2006){Ferrarese}, {C{\^o}t{\'e}}, {Jord{\'a}n},
  {Peng}, {Blakeslee}, {Piatek}, {Mei}, {Merritt}, {Milosavljevi{\'c}},
  {Tonry}, \& {West}}]{ferrarese06}
{Ferrarese}, L., {C{\^o}t{\'e}}, P., {Jord{\'a}n}, A., {et~al.} 2006, \apjs,
  164, 334

\bibitem[{{Fukugita} {et~al.}(1996){Fukugita}, {Ichikawa}, {Gunn}, {Doi},
  {Shimasaku}, \& {Schneider}}]{fukugita96}
{Fukugita}, M., {Ichikawa}, T., {Gunn}, J.~E., {et~al.} 1996, \aj, 111, 1748

\bibitem[{{Gonzalez} {et~al.}(2013){Gonzalez}, {Sivanandam}, {Zabludoff}, \&
  {Zaritsky}}]{gonzalez13}
{Gonzalez}, A.~H., {Sivanandam}, S., {Zabludoff}, A.~I., \& {Zaritsky}, D.
  2013, \apj, 778, 14

\bibitem[{{Gonzalez} {et~al.}(2007){Gonzalez}, {Zaritsky}, \&
  {Zabludoff}}]{gonzalez07}
{Gonzalez}, A.~H., {Zaritsky}, D., \& {Zabludoff}, A.~I. 2007, \apj, 666, 147

\bibitem[{{Guennou} {et~al.}(2012){Guennou}, {Adami}, {Da Rocha}, {Durret},
  {Ulmer}, {Allam}, {Basa}, {Benoist}, {Biviano}, {Clowe}, {Gavazzi},
  {Halliday}, {Ilbert}, {Johnston}, {Just}, {Kron}, {Kubo}, {Le Brun},
  {Marshall}, {Mazure}, {Murphy}, {Pereira}, {Raba{\c c}a}, {Rostagni},
  {Rudnick}, {Russeil}, {Schrabback}, {Slezak}, {Tucker}, \&
  {Zaritsky}}]{guennou12}
{Guennou}, L., {Adami}, C., {Da Rocha}, C., {et~al.} 2012, \aap, 537, A64

\bibitem[{{Jim{\'e}nez-Teja} \& {Dupke}(2016)}]{jimenez16}
{Jim{\'e}nez-Teja}, Y., \& {Dupke}, R. 2016, \apj, 820, 49

\bibitem[{{Kawara} {et~al.}(2017){Kawara}, {Matsuoka}, {Sano}, {Brandt},
  {Sameshima}, {Tsumura}, {Oyabu}, \& {Ienaka}}]{kawara17}
{Kawara}, K., {Matsuoka}, Y., {Sano}, K., {et~al.} 2017, \pasj,
  arXiv:1701.00885

\bibitem[{{Kitayama} {et~al.}(2009){Kitayama}, {Ito}, {Okada}, {Kaneda},
  {Takahashi}, {Ota}, {Onaka}, {Tajiri}, {Nagata}, \& {Yamada}}]{kitayama09}
{Kitayama}, T., {Ito}, Y., {Okada}, Y., {et~al.} 2009, \apj, 695, 1191

\bibitem[{{Koda} {et~al.}(2015){Koda}, {Yagi}, {Yamanoi}, \&
  {Komiyama}}]{koda15b}
{Koda}, J., {Yagi}, M., {Yamanoi}, H., \& {Komiyama}, Y. 2015, \apjl, 807, L2

\bibitem[{{Kodama} \& {Bower}(2001)}]{kodama01}
{Kodama}, T., \& {Bower}, R.~G. 2001, \mnras, 321, 18

\bibitem[{{Koyama} {et~al.}(2011){Koyama}, {Kodama}, {Nakata}, {Shimasaku}, \&
  {Okamura}}]{koyama11}
{Koyama}, Y., {Kodama}, T., {Nakata}, F., {Shimasaku}, K., \& {Okamura}, S.
  2011, \apj, 734, 66

\bibitem[{{Krick} \& {Bernstein}(2007)}]{krick07}
{Krick}, J.~E., \& {Bernstein}, R.~A. 2007, \aj, 134, 466

\bibitem[{{Kriek} {et~al.}(2009){Kriek}, {van Dokkum}, {Labb{\'e}}, {Franx},
  {Illingworth}, {Marchesini}, \& {Quadri}}]{kriek09}
{Kriek}, M., {van Dokkum}, P.~G., {Labb{\'e}}, I., {et~al.} 2009, \apj, 700,
  221

\bibitem[{{Larson} {et~al.}(1980){Larson}, {Tinsley}, \& {Caldwell}}]{larson80}
{Larson}, R.~B., {Tinsley}, B.~M., \& {Caldwell}, C.~N. 1980, \apj, 237, 692

\bibitem[{{Lee} {et~al.}(2017){Lee}, {Kang}, {Lee}, \& {Jang}}]{lee17}
{Lee}, M.~G., {Kang}, J., {Lee}, J.~H., \& {Jang}, I.~S. 2017, ArXiv e-prints,
  arXiv:1706.02521

\bibitem[{{Lotz} {et~al.}(2017){Lotz}, {Koekemoer}, {Coe}, {Grogin}, {Capak},
  {Mack}, {Anderson}, {Avila}, {Barker}, {Borncamp}, {Brammer}, {Durbin},
  {Gunning}, {Hilbert}, {Jenkner}, {Khandrika}, {Levay}, {Lucas}, {MacKenty},
  {Ogaz}, {Porterfield}, {Reid}, {Robberto}, {Royle}, {Smith},
  {Storrie-Lombardi}, {Sunnquist}, {Surace}, {Taylor}, {Williams}, {Bullock},
  {Dickinson}, {Finkelstein}, {Natarajan}, {Richard}, {Robertson}, {Tumlinson},
  {Zitrin}, {Flanagan}, {Sembach}, {Soifer}, \& {Mountain}}]{lotz17}
{Lotz}, J.~M., {Koekemoer}, A., {Coe}, D., {et~al.} 2017, \apj, 837, 97

\bibitem[{{Mantz} {et~al.}(2010){Mantz}, {Allen}, {Ebeling}, {Rapetti}, \&
  {Drlica-Wagner}}]{mantz10}
{Mantz}, A., {Allen}, S.~W., {Ebeling}, H., {Rapetti}, D., \& {Drlica-Wagner},
  A. 2010, \mnras, 406, 1773

\bibitem[{{Martel} {et~al.}(2012){Martel}, {Barai}, \& {Brito}}]{martel12}
{Martel}, H., {Barai}, P., \& {Brito}, W. 2012, \apj, 757, 48

\bibitem[{{McPartland} {et~al.}(2016){McPartland}, {Ebeling}, {Roediger}, \&
  {Blumenthal}}]{mcpartland16}
{McPartland}, C., {Ebeling}, H., {Roediger}, E., \& {Blumenthal}, K. 2016,
  \mnras, 455, 2994

\bibitem[{{Mihos} {et~al.}(2017){Mihos}, {Harding}, {Feldmeier}, {Rudick},
  {Janowiecki}, {Morrison}, {Slater}, \& {Watkins}}]{mihos17}
{Mihos}, J.~C., {Harding}, P., {Feldmeier}, J.~J., {et~al.} 2017, \apj, 834, 16

\bibitem[{{Montes} \& {Trujillo}(2014)}]{montes14}
{Montes}, M., \& {Trujillo}, I. 2014, \apj, 794, 137

\bibitem[{{Moore} {et~al.}(1996){Moore}, {Katz}, {Lake}, {Dressler}, \&
  {Oemler}}]{moore96}
{Moore}, B., {Katz}, N., {Lake}, G., {Dressler}, A., \& {Oemler}, A. 1996,
  \nat, 379, 613

\bibitem[{{Morishita} {et~al.}(2015){Morishita}, {Ichikawa}, {Noguchi},
  {Akiyama}, {Patel}, {Kajisawa}, \& {Obata}}]{morishita15}
{Morishita}, T., {Ichikawa}, T., {Noguchi}, M., {et~al.} 2015, \apj, 805, 34

\bibitem[{{Morishita} {et~al.}(2017){Morishita}, {Abramson}, {Treu}, {Vulcani},
  {Schmidt}, {Dressler}, {Poggianti}, {Malkan}, {Wang}, {Huang}, {Trenti},
  {Brada{\v c}}, \& {Hoag}}]{morishita17}
{Morishita}, T., {Abramson}, L.~E., {Treu}, T., {et~al.} 2017, \apj, 835, 254

\bibitem[{{Murante} {et~al.}(2007){Murante}, {Giovalli}, {Gerhard},
  {Arnaboldi}, {Borgani}, \& {Dolag}}]{murante07}
{Murante}, G., {Giovalli}, M., {Gerhard}, O., {et~al.} 2007, \mnras, 377, 2

\bibitem[{{Nipoti} {et~al.}(2003){Nipoti}, {Stiavelli}, {Ciotti}, {Treu}, \&
  {Rosati}}]{nipoti03}
{Nipoti}, C., {Stiavelli}, M., {Ciotti}, L., {Treu}, T., \& {Rosati}, P. 2003,
  \mnras, 344, 748

\bibitem[{{Nipoti} {et~al.}(2004){Nipoti}, {Treu}, {Ciotti}, \&
  {Stiavelli}}]{nipoti04}
{Nipoti}, C., {Treu}, T., {Ciotti}, L., \& {Stiavelli}, M. 2004, \mnras, 355,
  1119

\bibitem[{{Oke} \& {Gunn}(1983)}]{oke83}
{Oke}, J.~B., \& {Gunn}, J.~E. 1983, \apj, 266, 713

\bibitem[{{Ostriker} \& {Tremaine}(1975)}]{ostriker75}
{Ostriker}, J.~P., \& {Tremaine}, S.~D. 1975, \apjl, 202, L113

\bibitem[{{Peng} {et~al.}(2002){Peng}, {Ho}, {Impey}, \& {Rix}}]{peng02}
{Peng}, C.~Y., {Ho}, L.~C., {Impey}, C.~D., \& {Rix}, H.-W. 2002, \aj, 124, 266

\bibitem[{{Poggianti} {et~al.}(2009){Poggianti}, {Arag{\'o}n-Salamanca},
  {Zaritsky}, {De Lucia}, {Milvang-Jensen}, {Desai}, {Jablonka}, {Halliday},
  {Rudnick}, {Varela}, {Bamford}, {Best}, {Clowe}, {Noll}, {Saglia},
  {Pell{\'o}}, {Simard}, {von der Linden}, \& {White}}]{poggianti09}
{Poggianti}, B.~M., {Arag{\'o}n-Salamanca}, A., {Zaritsky}, D., {et~al.} 2009,
  \apj, 693, 112

\bibitem[{{Poggianti} {et~al.}(2016){Poggianti}, {Fasano}, {Omizzolo},
  {Gullieuszik}, {Bettoni}, {Moretti}, {Paccagnella}, {Jaff{\'e}}, {Vulcani},
  {Fritz}, {Couch}, \& {D'Onofrio}}]{poggianti16}
{Poggianti}, B.~M., {Fasano}, G., {Omizzolo}, A., {et~al.} 2016, \aj, 151, 78

\bibitem[{{Postman} {et~al.}(2012){Postman}, {Coe}, {Ben{\'{\i}}tez},
  {Bradley}, {Broadhurst}, {Donahue}, {Ford}, {Graur}, {Graves}, {Jouvel},
  {Koekemoer}, {Lemze}, {Medezinski}, {Molino}, {Moustakas}, {Ogaz}, {Riess},
  {Rodney}, {Rosati}, {Umetsu}, {Zheng}, {Zitrin}, {Bartelmann}, {Bouwens},
  {Czakon}, {Golwala}, {Host}, {Infante}, {Jha}, {Jimenez-Teja}, {Kelson},
  {Lahav}, {Lazkoz}, {Maoz}, {McCully}, {Melchior}, {Meneghetti}, {Merten},
  {Moustakas}, {Nonino}, {Patel}, {Reg{\"o}s}, {Sayers}, {Seitz}, \& {Van der
  Wel}}]{postman12}
{Postman}, M., {Coe}, D., {Ben{\'{\i}}tez}, N., {et~al.} 2012, \apjs, 199, 25

\bibitem[{{Presotto} {et~al.}(2014){Presotto}, {Girardi}, {Nonino}, {Mercurio},
  {Grillo}, {Rosati}, {Biviano}, {Annunziatella}, {Balestra}, {Cui},
  {Sartoris}, {Lemze}, {Ascaso}, {Moustakas}, {Ford}, {Fritz}, {Czoske},
  {Ettori}, {Kuchner}, {Lombardi}, {Maier}, {Medezinski}, {Molino},
  {Scodeggio}, {Strazzullo}, {Tozzi}, {Ziegler}, {Bartelmann}, {Benitez},
  {Bradley}, {Brescia}, {Broadhurst}, {Coe}, {Donahue}, {Gobat}, {Graves},
  {Kelson}, {Koekemoer}, {Melchior}, {Meneghetti}, {Merten}, {Moustakas},
  {Munari}, {Postman}, {Reg{\H o}s}, {Seitz}, {Umetsu}, {Zheng}, \&
  {Zitrin}}]{presotto14}
{Presotto}, V., {Girardi}, M., {Nonino}, M., {et~al.} 2014, \aap, 565, A126

\bibitem[{{Richard} {et~al.}(2010){Richard}, {Kneib}, {Limousin}, {Edge}, \&
  {Jullo}}]{richard10}
{Richard}, J., {Kneib}, J.-P., {Limousin}, M., {Edge}, A., \& {Jullo}, E. 2010,
  \mnras, 402, L44

\bibitem[{{Rudick} {et~al.}(2010){Rudick}, {Mihos}, {Harding}, {Feldmeier},
  {Janowiecki}, \& {Morrison}}]{rudick10}
{Rudick}, C.~S., {Mihos}, J.~C., {Harding}, P., {et~al.} 2010, \apj, 720, 569

\bibitem[{{Sayers} {et~al.}(2013){Sayers}, {Czakon}, {Mantz}, {Golwala},
  {Ameglio}, {Downes}, {Koch}, {Lin}, {Maughan}, {Molnar}, {Moustakas},
  {Mroczkowski}, {Pierpaoli}, {Shitanishi}, {Siegel}, {Umetsu}, \& {Van der
  Pyl}}]{sayers13}
{Sayers}, J., {Czakon}, N.~G., {Mantz}, A., {et~al.} 2013, \apj, 768, 177

\bibitem[{{Toledo} {et~al.}(2011){Toledo}, {Melnick}, {Selman}, {Quintana},
  {Giraud}, \& {Zelaya}}]{toledo11}
{Toledo}, I., {Melnick}, J., {Selman}, F., {et~al.} 2011, \mnras, 414, 602

\bibitem[{{Treu} {et~al.}(2003){Treu}, {Ellis}, {Kneib}, {Dressler}, {Smail},
  {Czoske}, {Oemler}, \& {Natarajan}}]{treu03}
{Treu}, T., {Ellis}, R.~S., {Kneib}, J.-P., {et~al.} 2003, \apj, 591, 53

\bibitem[{{Treu} {et~al.}(2005){Treu}, {Ellis}, {Liao}, {van Dokkum}, {Tozzi},
  {Coil}, {Newman}, {Cooper}, \& {Davis}}]{treu05}
{Treu}, T., {Ellis}, R.~S., {Liao}, T.~X., {et~al.} 2005, \apj, 633, 174

\bibitem[{{Tsumura} {et~al.}(2013){Tsumura}, {Matsumoto}, {Matsuura}, {Sakon},
  {Tanaka}, \& {Wada}}]{tsumura13}
{Tsumura}, K., {Matsumoto}, T., {Matsuura}, S., {et~al.} 2013, \pasj, 65, 120

\bibitem[{{van Dokkum} {et~al.}(2015){van Dokkum}, {Abraham}, {Merritt},
  {Zhang}, {Geha}, \& {Conroy}}]{vandokkum15a}
{van Dokkum}, P.~G., {Abraham}, R., {Merritt}, A., {et~al.} 2015, \apjl, 798,
  L45

\bibitem[{{Vijayaraghavan} \& {Ricker}(2013)}]{vija13}
{Vijayaraghavan}, R., \& {Ricker}, P.~M. 2013, \mnras, 435, 2713

\bibitem[{{Vulcani} {et~al.}(2016{\natexlab{a}}){Vulcani}, {Treu}, {Schmidt},
  {Morishita}, {Dressler}, {Poggianti}, {Abramson}, {Brada{\v c}}, {Brammer},
  {Hoag}, {Malkan}, {Pentericci}, \& {Trenti}}]{vulcani16a}
{Vulcani}, B., {Treu}, T., {Schmidt}, K.~B., {et~al.} 2016{\natexlab{a}}, ArXiv
  e-prints, arXiv:1610.04621

\bibitem[{{Vulcani} {et~al.}(2016{\natexlab{b}}){Vulcani}, {Treu}, {Nipoti},
  {Schmidt}, {Dressler}, {Morshita}, {Poggianti}, {Malkan}, {Hoag}, {Brada{\v
  c}}, {Abramson}, {Trenti}, {Pentericci}, {von der Linden}, {Morris}, \&
  {Wang}}]{vulcani16b}
{Vulcani}, B., {Treu}, T., {Nipoti}, C., {et~al.} 2016{\natexlab{b}}, ArXiv
  e-prints, arXiv:1610.04615

\bibitem[{{West} {et~al.}(2017){West}, {de Propris}, {Bremer}, \&
  {Phillipps}}]{west17}
{West}, M.~J., {de Propris}, R., {Bremer}, M.~N., \& {Phillipps}, S. 2017,
  Nature Astronomy, 1, 0157

\bibitem[{{Williams} {et~al.}(2007){Williams}, {Ciardullo}, {Durrell},
  {Vinciguerra}, {Feldmeier}, {Jacoby}, {Sigurdsson}, {von Hippel}, {Ferguson},
  {Tanvir}, {Arnaboldi}, {Gerhard}, {Aguerri}, \& {Freeman}}]{williams07}
{Williams}, B.~F., {Ciardullo}, R., {Durrell}, P.~R., {et~al.} 2007, \apj, 656,
  756

\bibitem[{{Yagi} {et~al.}(2013){Yagi}, {Gu}, {Fujita}, {Nakazawa}, {Akahori},
  {Hattori}, {Yoshida}, \& {Makishima}}]{yagi13}
{Yagi}, M., {Gu}, L., {Fujita}, Y., {et~al.} 2013, \apj, 778, 91

\bibitem[{{Yagi} {et~al.}(2016){Yagi}, {Koda}, {Komiyama}, \&
  {Yamanoi}}]{yagi16}
{Yagi}, M., {Koda}, J., {Komiyama}, Y., \& {Yamanoi}, H. 2016, \apjs, 225, 11

\bibitem[{{Yoshida} {et~al.}(2002){Yoshida}, {Yagi}, {Okamura}, {Aoki},
  {Ohyama}, {Komiyama}, {Yasuda}, {Iye}, {Kashikawa}, {Doi}, {Furusawa},
  {Hamabe}, {Kimura}, {Miyazaki}, {Miyazaki}, {Nakata}, {Ouchi}, {Sekiguchi},
  {Shimasaku}, \& {Ohtani}}]{yoshida02}
{Yoshida}, M., {Yagi}, M., {Okamura}, S., {et~al.} 2002, \apj, 567, 118

\bibitem[{{Zabludoff} \& {Mulchaey}(1998)}]{zabludoff98}
{Zabludoff}, A.~I., \& {Mulchaey}, J.~S. 1998, \apj, 496, 39

\bibitem[{{Zaritsky} {et~al.}(2004){Zaritsky}, {Gonzalez}, \&
  {Zabludoff}}]{zaritsky04}
{Zaritsky}, D., {Gonzalez}, A.~H., \& {Zabludoff}, A.~I. 2004, \apjl, 613, L93

\bibitem[{{Zibetti} {et~al.}(2005){Zibetti}, {White}, {Schneider}, \&
  {Brinkmann}}]{zibetti05}
{Zibetti}, S., {White}, S.~D.~M., {Schneider}, D.~P., \& {Brinkmann}, J. 2005,
  \mnras, 358, 949

\bibitem[{{Zwicky}(1937)}]{zwicky37}
{Zwicky}, F. 1937, \apj, 86, 217

\end{thebibliography}

\end{document}